\documentclass[]{revtex4}

\usepackage{graphics} % for pdf, bitmapped graphics files
\usepackage[pdftex]{graphicx}
\usepackage{epstopdf}
\usepackage[cmex10]{amsmath}
\usepackage{amssymb}  % assumes amsmath package installed
\usepackage[colorlinks,linkcolor=blue,pdfauthor=Alexander \ Scheinker,citecolor=red]{hyperref}
\usepackage{enumitem}
\usepackage{comment} 
\usepackage{pdfpages}

\begin{document}

%\includepdf[pages={1}]{ACM4PA_cover_vFinal.pdf}

%Title of paper
\title{Advanced Control Methods for Particle Accelerators (ACM4PA) 2019}\thanks{This workshop was hosted by Los Alamos National Laboratory at Hotel Santa Fe, in Santa Fe, NM, USA, from Tuesday, August 19 to Thursday, August 22, 2019.}

\author{Alexander~Scheinker}
\email[]{ascheink@lanl.gov}
\affiliation{Los Alamos National Laboratory, Los Alamos, NM, 87544, USA}
\author{Claudio~Emma}
\email[]{cemma@slac.stanford.edu}
\author{Auralee~L~Edelen}
\email[]{edelen@slac.stanford.edu}
\author{Spencer~Gessner} 
\email[]{sgess@slac.stanford.edu}
\affiliation{SLAC National Accelerator Laboratory, Menlo Park, CA, 94025, USA}

\date{\today}

\begin{abstract}
Los Alamos is currently developing novel particle accelerator controls and diagnostics algorithms to enable higher quality beams with lower beam losses than is currently possible. The purpose of this workshop was to consider tuning and optimization challenges of a wide range of particle accelerators including linear proton accelerators such as the Los Alamos Neutron Science Center (LANSCE), rings such as the Advanced Photon Source (APS) synchrotron, free electron lasers (FEL) such as the Linac Coherent Light Source (LCLS) and LCLS-II, the European X-ray Free Electron Laser (EuXFEL), the Swiss FEL, and the planned MaRIE FEL, and plasma wake-field accelerators such as FACET, FACET-II, and AWAKE at CERN. One major challenge is an the ability to quickly create very high quality, extremely intense, custom current and energy profile beams while working with limited real time non-invasive diagnostics and utilizing time-varying uncertain initial beam distributions and accelerator components. Currently, a few individual accelerator labs have been developing and applying their own diagnostics tools and custom control and ML algorithms for automated machine tuning and optimization. The goal of this workshop was to bring together a group of accelerator physicists and accelerator related control and ML experts in order to define which controls and diagnostics would be most useful for existing and future accelerators and to create a plan for developing a new family of algorithms that can be shared and maintained by the community. \\
\\
{\noindent}{\bf Organizing Committee} \\
Alexander Scheinker\hfill LANL \\
Spencer Gessner\hfill CERN \\
Claudio Emma\hfill SLAC \\
Auralee Edelen\hfill SLAC \\

{\noindent}{\bf Invited Speakers} \\
Sam Barber\hfill LBNL \\
Kip Bishofberger\hfill LANL \\
Joseph Duris\hfill SLAC \\
Sandy Easton\hfill CERN \\
Jonathan Edelen\hfill RadiaSoft \\
Adi Hanuka\hfill SLAC \\
John Lewellen\hfill LANL \\
Jerry Ling\hfill UCSB \\
Quinn Marksteiner\hfill LANL \\
Brendan O'Shea\hfill SLAC \\
Tor Raubenheimer\hfill SLAC \\
Samuel Schoenholz\hfill GOOGLE \\
Anna Solopova Shabalina\hfill JLAB \\
Yine Sun\hfill ANL \\
Faya Wang\hfill SLAC \\

{\noindent}{\bf Workshop Website} \\
\url{http://www.cvent.com/d/dyq71b}

\end{abstract}

\maketitle

\section{2019 ACM4PA}

\begin{figure*}[!th]
\centering
{\includegraphics[width=0.8\textwidth]{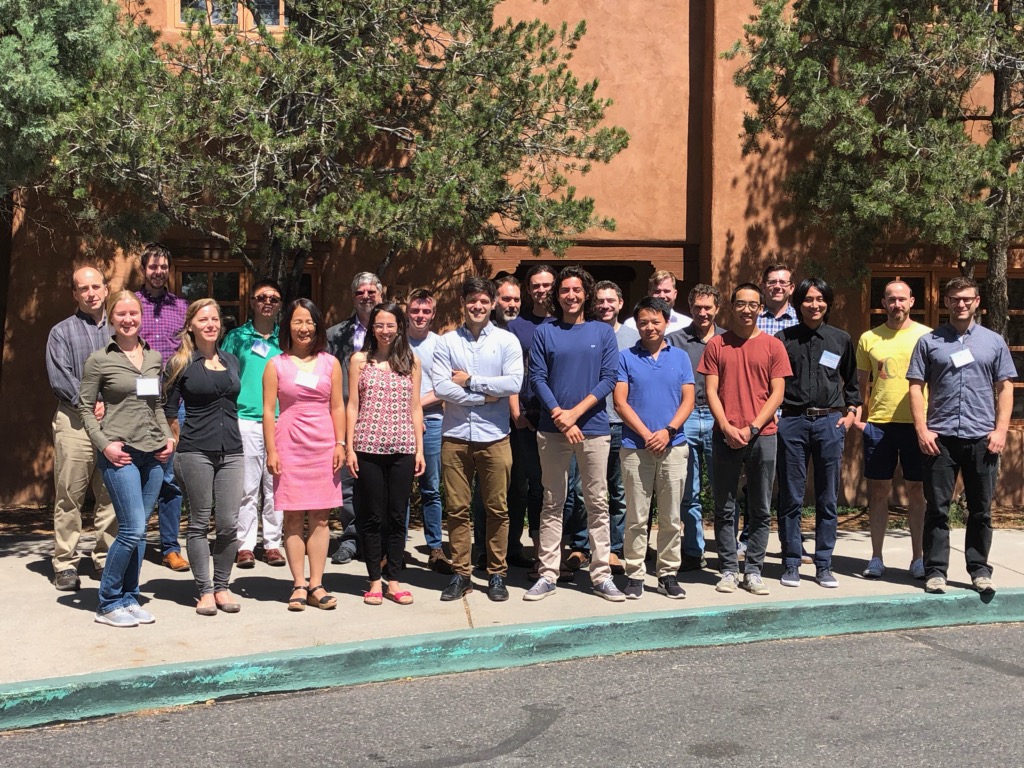}}
\caption{Group photo of participants at the 2019 ACM4PA.}
\label{fig:workshop}
\end{figure*}

\pagebreak

\section{Schedule}

\begin{figure*}[!th]
\centering
{\includegraphics[width=0.8\textwidth]{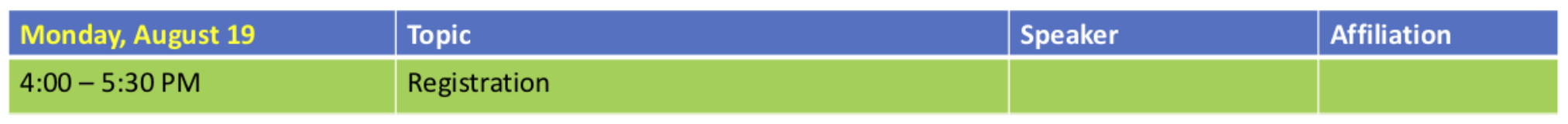}} \\
{\includegraphics[width=0.8\textwidth]{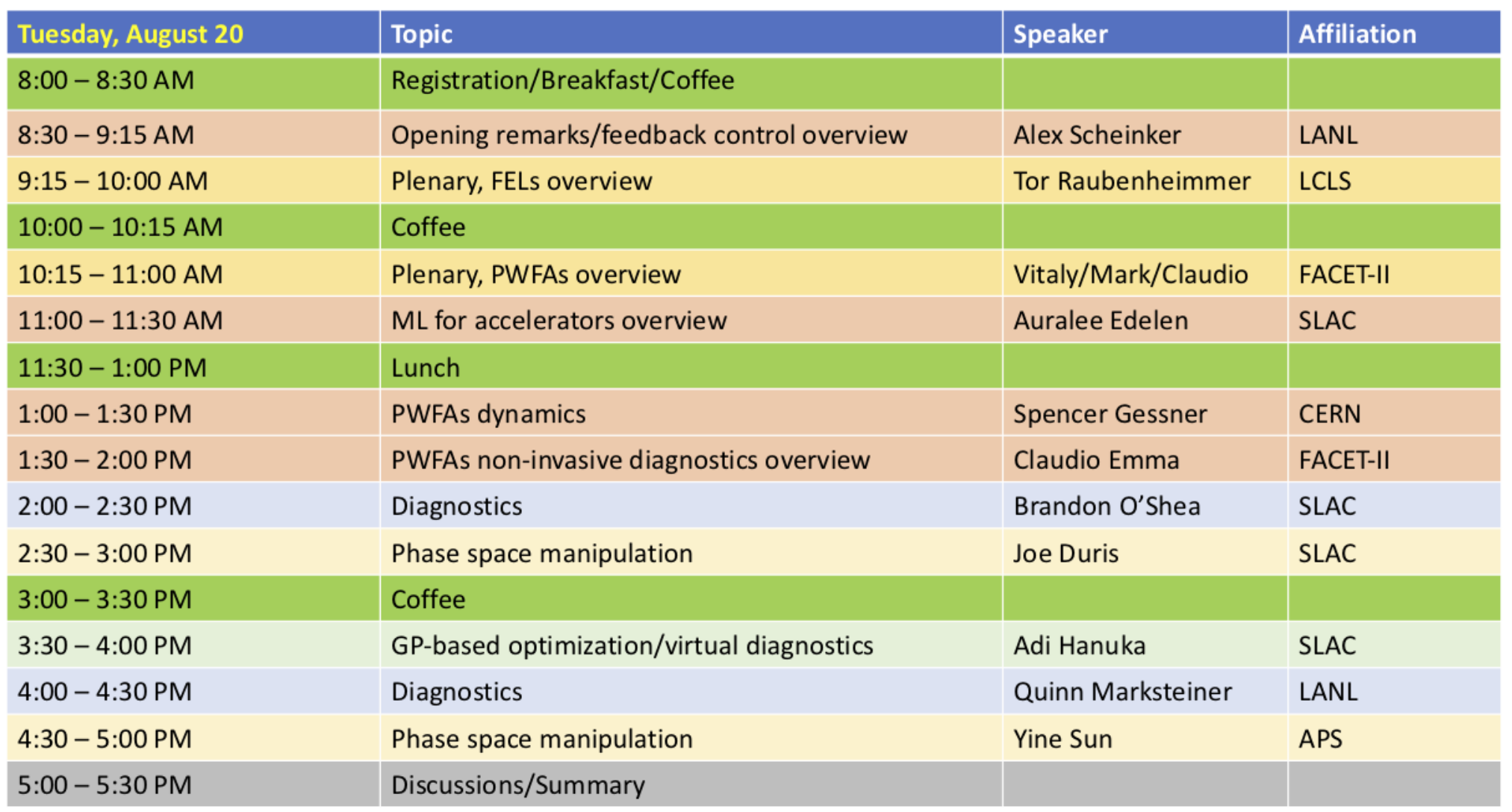}} \\
{\includegraphics[width=0.8\textwidth]{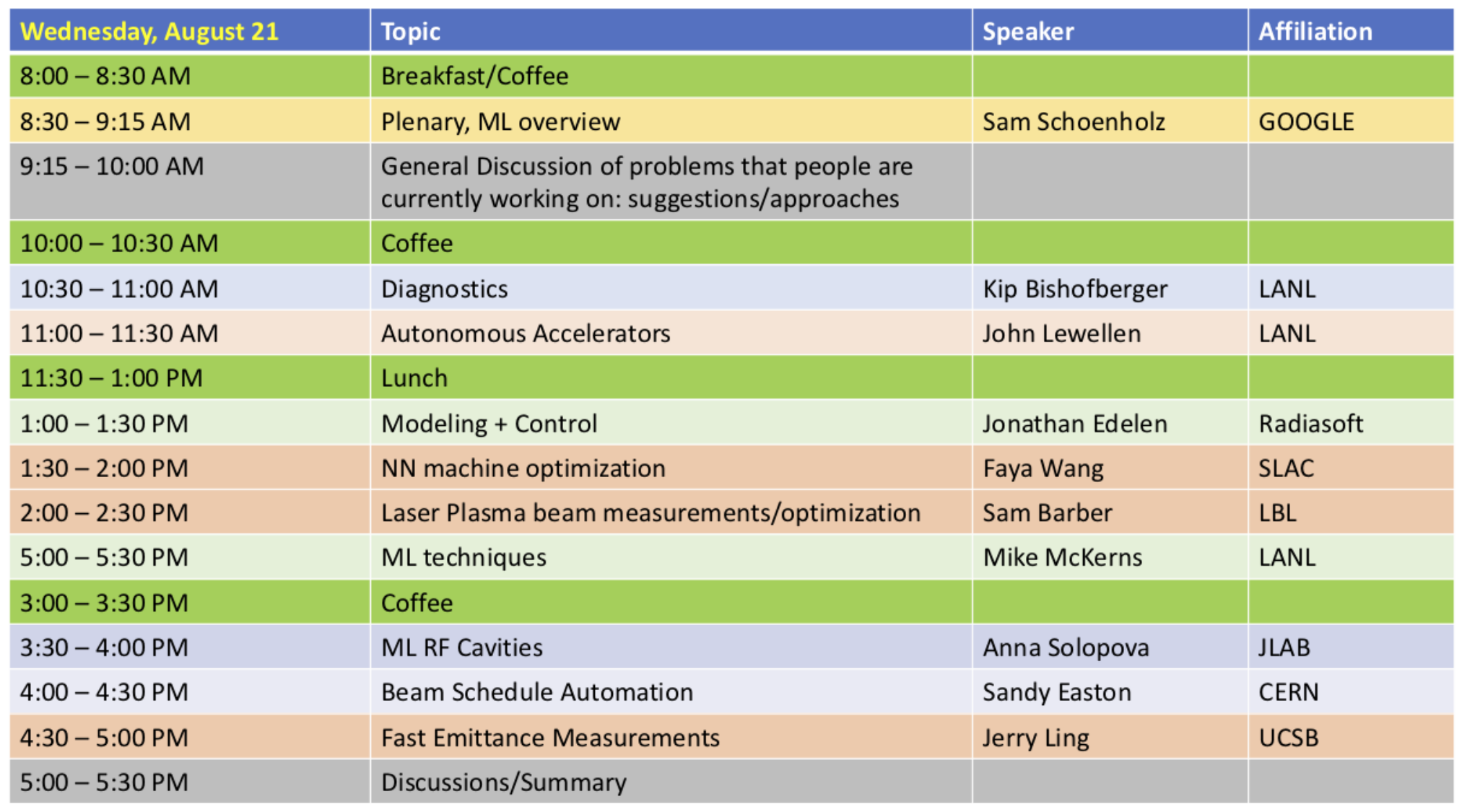}} \\
{\includegraphics[width=0.8\textwidth]{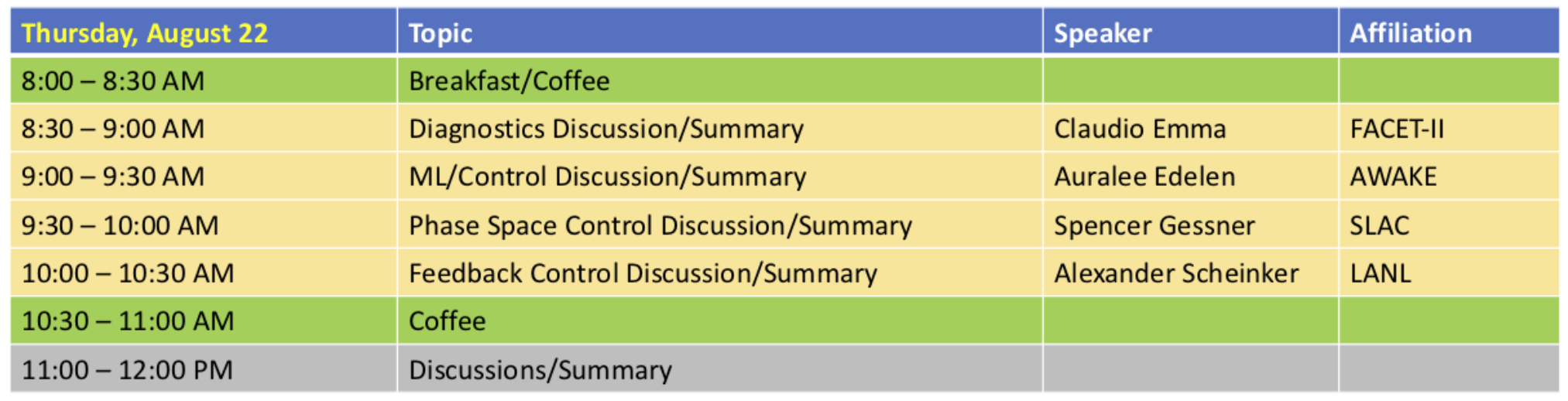}}
\caption{Schedule of talks and discussions at the 2019 ACM4PA.}
\label{fig:schedule}
\end{figure*}

\pagebreak

%%%%%%%%%%%%%%%%%%%%%%%%%%%%%%%%%%%%%%%%%%%%%%%%%%%%%%%%%%%%%%%%%%%%%%%%%%%%%%%%%
%%%%%%%%%%%%%%%%%%%%%%%%%%%%%%%%%%%%%%%%%%%%%%%%%%%%%%%%%%%%%%%%%%%%%%%%%%%%%%%%%
%%%%%%%%%%%%%%%%%%%%%%%%%%%%%%%%%%%%%%%%%%%%%%%%%%%%%%%%%%%%%%%%%%%%%%%%%%%%%%%%%
\section{Motivation}
%%%%%%%%%%%%%%%%%%%%%%%%%%%%%%%%%%%%%%%%%%%%%%%%%%%%%%%%%%%%%%%%%%%%%%%%%%%%%%%%%
%%%%%%%%%%%%%%%%%%%%%%%%%%%%%%%%%%%%%%%%%%%%%%%%%%%%%%%%%%%%%%%%%%%%%%%%%%%%%%%%%
%%%%%%%%%%%%%%%%%%%%%%%%%%%%%%%%%%%%%%%%%%%%%%%%%%%%%%%%%%%%%%%%%%%%%%%%%%%%%%%%%

\begin{figure*}[!th]
\centering
{\includegraphics[width=1.0\textwidth]{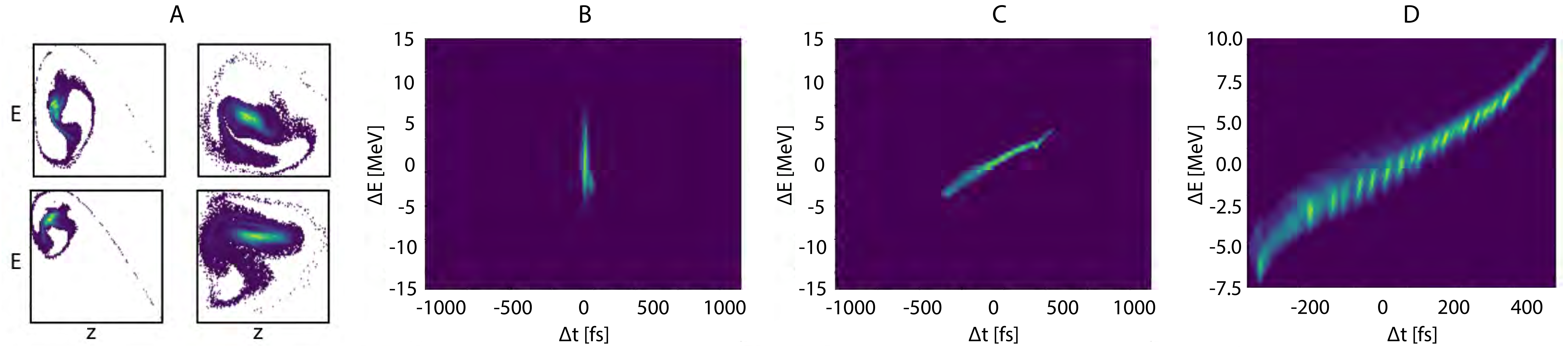}}
\caption{Simulation of a proton bunch passing through LANSCE is shown in (A). Nonlinear space charge forces twist the beam, create halos and tails of the bunch particle distribution in energy vs time phase space. Results of the first stage of bunch compression at the EuXFEL is shown in (B-D). The bunch in (B) is nicely compressed, creating a short, intense beam. The bunch in (C) is not as tightly compressed as the one in (B), and is suffering from microbunch instabilities where local density fluctuations naturally occurring in the bunch are amplified by the compression process, as seen in more detail in (D).}
\label{fig:TCAV}
\end{figure*}

Precise control of the longitudinal phase space (LPS) (time vs energy: bunch length, current profile, energy spread) of charged particle bunches is required for all particle accelerators. The following outlines some accelerator capabilities and challenges (from \cite{ref-ES-NN}):

{\it Free electron lasers (FEL) are incredibly powerful scientific tools for studying physics at previously inaccessible length (nanometers) and time (femtoseconds (fs)) scales for high energy physics, biology, chemistry, material science, and accelerator physics experiments \cite{ref-FELs,ref-FELs2,ref-FELs3}. FELs can produce extremely short ($\sim$fs) coherent X-ray bursts with tunable wavelength which are many orders of magnitude brighter than traditional sources such as synchrotrons. For example, the Linac Coherent Light Source (LCLS) FEL, the first hard X-ray FEL, provides users with photon energies ranging from 0.27 keV to 12 keV based on electron bunches with energies from 2.5 GeV to 17 GeV. Operating electron bunch charge can range from 20 pC to 300 pC and the bunch duration from 3 fs to 500 fs to suit experimental needs \cite{ref-LCLS,ref-LCLS2,ref-LCLS3}. The European X-Ray FEL (EuXFEL), one of the newest and most advanced FELs in the world, is capable of producing 27000 pulses of bright, coherent light per second utilizing electron bunches with energies of up to 17.5 GeV, with charges ranges from 0.02 to 1 nC per bunch, and photon energies from 0.26 keV up to 25 keV \cite{ref-EuXFEL,ref-EuXFEL2,ref-EuXFEL3}. Precise control of bunch lengths, current profiles, and energy spreads of increasingly shorter electron beams at femtosecond resolution is extremely important and challenging for both the LCLS and the Eu-XFEL \cite{ref-LCLS-Tuning,ref-fs}.

The extremely bright and short X-ray bursts that make FELs such powerful instruments also makes them incredibly challenging to control. High power FELs are driven by few kilometer long high power particle accelerators composed of thousands of interacting electromagnetic components including radio frequency (RF) accelerating cavities and magnets. The performance of all of these components is susceptible to drift, e.g. such as thermal drifts. A major challenge is the FEL lasing process itself, self-amplified spontaneous emission (SASE) is stochastic in nature and extremely sensitive to the beam's initial conditions including charge density and energy spread, therefore there is a large variance in the output power of FEL light even when all machine set points are fixed and properly timed because of uncertainty in and time variation of the electron distribution coming off of the photo cathode and entering the accelerator. Therefore, the output power of an FEL is a highly nonlinear, time-varying, noisy and analytically unknown function of all of the FELs thousands of components. Traditional model-based approaches are severely limited by such uncertainties and time variation of both the accelerated beam's phase space distribution and the accelerator's components as well as misalignments, thermal cycling, and collective effects such as space charge forces, wakefields, and coherent synchrotron radiation emitted by extremely short high current bunches.

One example of such difficulties is the process of reconfiguring the LCLS to a low charge mode to provide 3 fs bunches, a process which may require many hours of expert hands on tuning. These difficulties will only grow for future facilities, such as the LCLS-II \cite{ref-LCLS-II} for complex schemes such as multi-color operation \cite{ref-LCLS2color}, multi-stage amplification \cite{ref-LutmanFreshSlice} or self-seeding \cite{ref-LCLS3,ref-Amman}. Plasma wakefield accelerators (PWFA) are another class of particle accelerators which require extremely intense, high current and sometimes extremely short charged particle bunches with complex beam dynamics and phase space manipulations \cite{ref-FACET}. For example, the facility for advanced accelerator experimental tests (FACET-II) is being designed to provide custom tailored current profiles for various experiments with bunch lengths as low as 1 or 2 fs \cite{ref-FACET-II2,ref-FACET-II}.
\it}

One illustrative problem is that of creating tightly packed charged particle bunches and maintaining these bunches while accelerating them. Accelerators utilize radio frequency (RF) resonant cavities to create large (tens of MV/m) electric field gradients with energy gain proportional to $V \cos \left ( \phi \right )$ with $\phi$ the phase of the electromagnetic wave at the time of a particle's arrival. Non-zero $\phi$ creates a precise energy deviation along the electron bunch. For a linear proton accelerator such as LANSCE, energy difference translates to significant velocity difference causing beam bunching. For highly relativistic electron accelerators, such as PWFAs and FELs, the velocity difference is insignificant (v $\approx$ c), so a beam is passed through a magnetic chicane translating energy difference to path length difference, compressing the bunch by orders of magnitude. For an RF field of frequency $\omega_{\mathrm{rf}}$, the phase $\phi$ relative to the RF at position offset $\Delta z_0$ is given by $\phi = -\omega_{\mathrm{rf}} \Delta z_0 /c$. For small $\phi$, for an initial bunch length of $\sigma_{z0}$ has approximate final bunch length
\begin{equation}
	\sigma_{zf} \approx \sqrt{ \left ( 1 + R_{56} \frac{eV_{\mathrm{rf}}\omega_{\mathrm{rf}}}{cE} \right )^2 \sigma^2_{z0} + R^2_{56}\sigma^2_{\Delta E_0}}, \quad
	R_{56}(z) = \int_{z_0}^{z}\frac{R_{16}(z')}{\rho(z')}dz', \label{compress}
\end{equation}
where $\sigma_{\Delta E_0}$ is initial beam energy spread, $R_{16}$ is transverse displacement resulting from energy offset in a dispersive region of the accelerator, and $\rho(z)$ the local radius of curvature for a beam passing through bending magnets down the length of the accelerator (z). In practice bunch compression is a much more complex process than (\ref{compress}). Intense particle bunches undergo collective effects such as space charge forces resulting in collective phenomenon such as halo formation and micro-bunch instability creating local density fluctuations. Precise setup and control of bunch compression requires non-invasive diagnostics to ensure that beam quality is preserved. Figure \ref{fig:TCAV} shows LPS simulations of proton bunches in LANSCE and TCAV measurements of the LPS of electron bunches in the EuXFEL after the first bunch compressor. 

\begin{figure*}[!th]
\centering
{\includegraphics[width=0.8\textwidth]{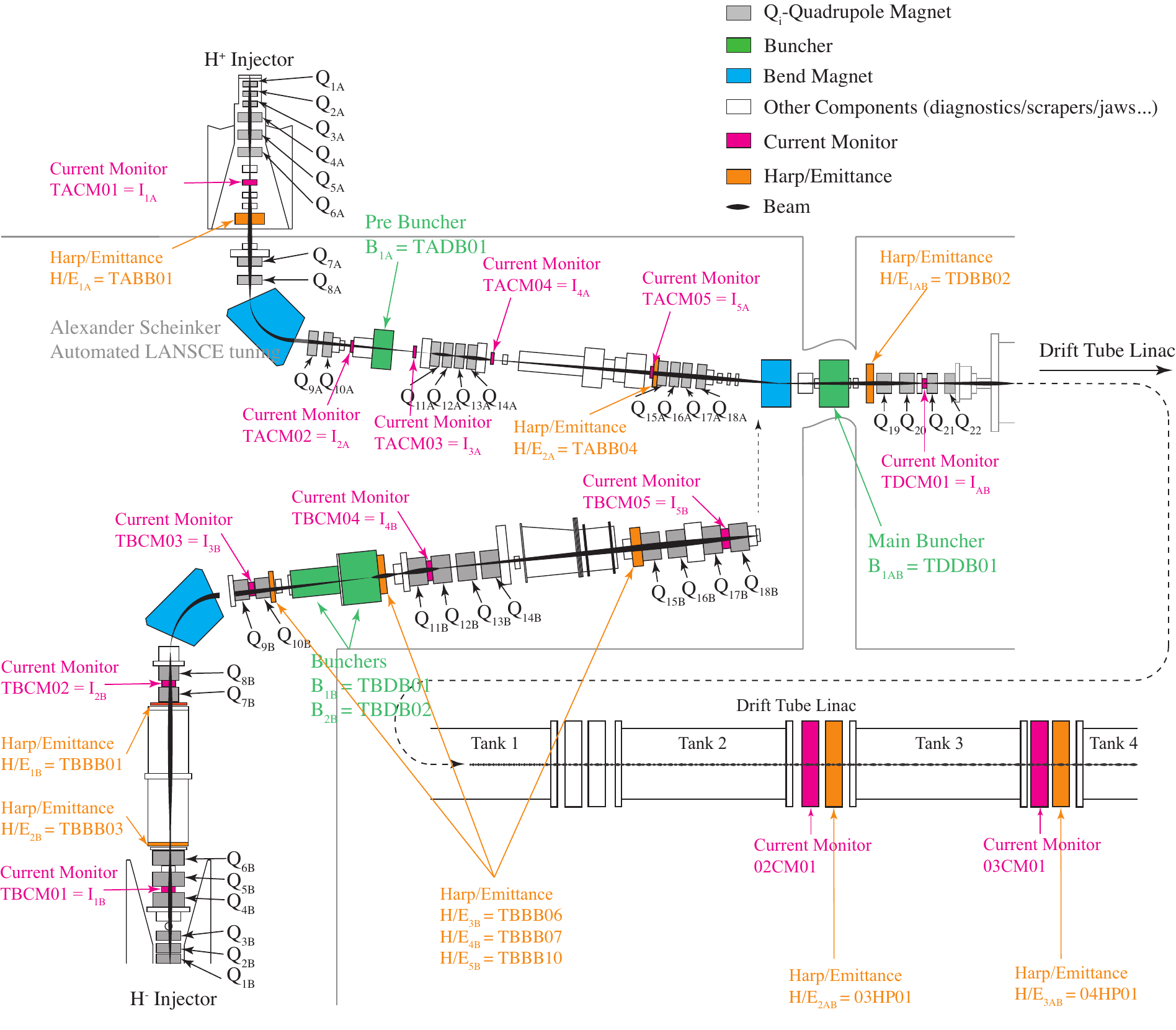}}
\caption{Overview of the injector and drift tube linac sections of the LANSCE proton linac at Los Alamos National Laboratory.}
\label{fig:LANSCE_layout}
\end{figure*}

Linear proton machines, such as the Los Alamos Neutron Science Center (LANSCE), also face tuning challenges because of highly space charge dominated beams, limited diagnostics, and the complexity of many coupled components, as shown in Figure \ref{fig:LANSCE_layout}. 

\vfill

\pagebreak

%%%%%%%%%%%%%%%%%%%%%%%%%%%%%%%%%%%%%%%%%%%%%%%%%%%%%%%%%%%%%%%%%%%%%%%%%%%%%%%%%
%%%%%%%%%%%%%%%%%%%%%%%%%%%%%%%%%%%%%%%%%%%%%%%%%%%%%%%%%%%%%%%%%%%%%%%%%%%%%%%%%
%%%%%%%%%%%%%%%%%%%%%%%%%%%%%%%%%%%%%%%%%%%%%%%%%%%%%%%%%%%%%%%%%%%%%%%%%%%%%%%%%
\section{Summary}
%%%%%%%%%%%%%%%%%%%%%%%%%%%%%%%%%%%%%%%%%%%%%%%%%%%%%%%%%%%%%%%%%%%%%%%%%%%%%%%%%
%%%%%%%%%%%%%%%%%%%%%%%%%%%%%%%%%%%%%%%%%%%%%%%%%%%%%%%%%%%%%%%%%%%%%%%%%%%%%%%%%
%%%%%%%%%%%%%%%%%%%%%%%%%%%%%%%%%%%%%%%%%%%%%%%%%%%%%%%%%%%%%%%%%%%%%%%%%%%%%%%%%

{\noindent}New algorithms for non-invasive diagnostics and control based on adaptive feedback (AF) and machine learning (ML) methods can greatly benefit existing accelerators as well as enable future facilities such as the planned MaRIE, in particular, new adaptive machine learning methods combining AF and ML for complex time-varying systems \cite{ref-ES-NN}. 
\\

{\noindent}{{\bf Diagnostics}
\\
New diagnostics requirements especially emphasize the need for non-invasive diagnostics that can provide real time information about extremely short and intense charged particle bunches. Many new approaches for non-invasive diagnostics that are being discussed are planning to use the spectrum of light emitted from accelerating electrons beams in order to attempt to reconstruct their current profiles. One such approach is to pass a high energy electron beam through a gentle undulator and then utilize the off-axis radiation to reconstruct the electron density. This has not yet been demonstrated, especially the unique iterative phase retrieval problem needs to be considered carefully. Another technique being considered is optical diffraction radiation created when a beam grazes a metal edge, giving a 1D projection of a slice of a beam's 2D phase space. Numerical methods show promise in using several independent 1D projections to reconstruct a 2D phase space, giving transverse beam diagnostics. There are also computational approaches being considered for diagnostics, one of which is to attempt to train a neural network to map accelerator parameter settings to longitudinal phase space (LPS) distributions, this approach has been demonstrated in simulation and at the LCLS \cite{ref-ML4}. Another approach is to adaptively tune accelerator models based on information non-invasively gathered about the beam to get an actual predictive match between model and machine. A preliminary version of such an approach was demonstrated at FACET where a non-invasive energy spread spectrum was used to adaptively tune a model in real time to give predictions of the LPS of the electron beam which was confirmed with TCAV measurements \cite{ref-ES-FACET}. This method was demonstrated to be extremely robust to time-variation of the system, further studies are being planned at LANSCE and FACET-II.
\\

{\noindent}{{\bf Optimization for learning}
\\
Advanced algorithms have a role to play both in optimization experiments and learning underlying physics. For example, if an optimizer is given the freedom to adjust many parameters at once in order to, for example, maximize the energy gain while minimizing the energy spread of the beam exiting a plasma wakefield accelerator, the optimizer may find an unexpected and not analytically predicted configuration which achieves better performance than model-based methods. Studying this configuration will then teach us either about the machine and what components are not operating in an ideal way and therefore requiring non-ideal component settings, or, more interestingly it could possibly find a new configuration based on physical affects that were not considered in simplified models, which, when understood may lead to the discovery of new methods and approaches to the problem being studied.
\\

{\noindent}{\bf Differentiable programing}
\\
Many machine learning codes, such as the Tensorflow environment created by Google for the development of neural networks, are differentiable environments. In these codes, once equations are written, they are implemented as directed graphs with analytically determined derivatives computed for each graph segment, resulting in input-output maps which are automatically differentiable. One very useful feature of this that should be taken advantage of is, for example, the creation of a neural network to learn the mapping between various accelerator components and beam characteristics such as emittance or energy spread. This mapping will then be differentiable and therefore it will be possible to directly differentiate, for example, energy spread versus a particular magnet settings, for a very large complex system based only on data gathered from a real machine. This approach would then provide a differentiable map which would be very useful for everything from parameter sensitivity studies to optimization.
\\

{\noindent}{{\bf Time-varying systems (distribution shifts)}
\\
Although ML techniques such as deep convolutional neural networks (CNN) have recently become popular due to a growth in computational power, their implementation by the accelerator community so far has been limited to a few test cases. One of the main challenges of applying ML techniques to accelerator applications stems from the fact that the performance of model/data-based ML techniques, such as CNNs, suffers when the system for which they have been trained quickly changes with time. Accelerators suffer from time-varying distribution shifts of, for example, the phase space of the particle bunches entering the accelerators, and also of the performance of the hundreds of magnets and RF systems, which drift with time. Approaches for handling distributions shifts include: 
\\
\\
{\bf 1). Adaptive ML:} To supplement ML approaches, local, model-independent feedback algorithms exist in the feedback control theory community, that can be extremely helpful for optimizing and tuning noisy complex systems and to adapt to time-varying features and distributions \cite{ref-ES,ref-Sch-Sch,ref-ES-book,ref-Sch-Sch-2,ref-ES-LANSCE-phase,ref-ES-LANSCE,ref-ES-SPEAR3,ref-ES-EuXFEL}. One application of this was demonstrated at the LCLS for automatically tuning the FEL to achieve desired longitudinal phase distributions \cite{ref-ES-NN}.
\\
{\bf 2). Gaussian processes (GP):} When statistically significant amounts of data can be gathered relating a set of parameters to each other and to a particular objective function (such as quadrupole magnets to X-ray pulse energy), GP methods may be used to give predictions for component settings as well as uncertainty bounds. Such methods are being developed and tested at the LCLS greatly outperforming standard methods such as simplex and robust conjugate direction search (RCDS) \cite{ref-LCLS-Gauss,ref-ML-Adi}. 
\\
{\bf 3). Re-training:} Whether using a particular data set collected from an accelerator over a certain period of time or if learning based on simulation data, it may be possible to re-train only a small final section of a pre-trained NN to give better predictions relative to the latest collected data in order to re-learn continuously as things vary with time. However, for very large numbers of parameters and large networks this may be infeasible as it would take too much data and too long to re-train relative to how fast the system is actually changing. Currently such methods are used for static problems, such as image identification and classification. 
\\
{\bf 4). Domain transfer:} Another approach for NNs is to initially train based on a very large amount of simulation or experimental data and then quickly train a much smaller network based on new data, a new network which will map readings to their simulation-based counterparts to be fed into the NN. So, for example, if the NN was trained to map a particular magnet setting to a phase space distribution, but that relationship later changed, a smaller domain transfer NN, such as a U-net, could be trained to map the new magnet setting to the old simulation-based setting, to be fed into the NN to predict the new phase space. Such a domain transfer NN would have to be continuously re-trained on new data. Again, this may become infeasible for a NN handling a large number of parameters because the retraining may take too long or would require too large of a manual grid-scan of the parameter space. Currently such methods are used for static problems, such as image identification and classification. One particularly interesting recent use of this method, which has great potential for particle accelerator applications, was to utilize large amounts of simulation data to initially train a CNN to map diffraction images to crystal orientations and then apply both re-training and domain transfer to tune the CNN or to adjust its inputs for it to accurately work on actual experimental data which was not exactly matched by the simulations \cite{ref-ML-retrain}.
\\

{\noindent}{{\bf Workshop-based Collaboration}\\
One direct result of this workshop was the establishment of a collaboration between Los Alamos and CERN. CERN invited A. Scheinker to perform automatic optimization and tuning studies on the electron beam line of the CERN AWAKE plasma wakefield accelerator experiment. 

Working together with CERN scientists for one week they were able to implement a LANL-developed adaptive extremum seeking (ES) tuning algorithm on the CERN machine. In this work, two ES algorithms were run simultaneously at two different time scales. One ES algorithm, ES$_1$ slowly tuned 5 parameters: 2 solenoid currents and 3 quadrupole magnets to adjust the transverse phase space of the electron bunch, in order to minimize the size of the electron beam at the end of the beam line. However, while this algorithm was running, it not only affected the transverse phase space, but also the longitudinal: the energy of the beam and therefore its orbit was being modified and the beam was moving away from its design orbit and off of the measurement screen at the end of the beam line. A second ES algorithm, ES$_2$ was also run, operating $\sim3\times$ faster than ES$_1$. The second algorithm, ES$_2$ adjusted 9 steering magnets based on 9 beam position monitor readings to continuously maintain the desired orbit, while ES$_1$ worked to minimize the beam size. The overall result was a beam size approximately $2\times$ smaller than what was previously achieved. The experimental results are shown in Figure \ref{fig:CERN_emittance} and are being prepared for a journal publication.

\begin{figure}[!th]
\centering
{\includegraphics[width=0.345\textwidth]{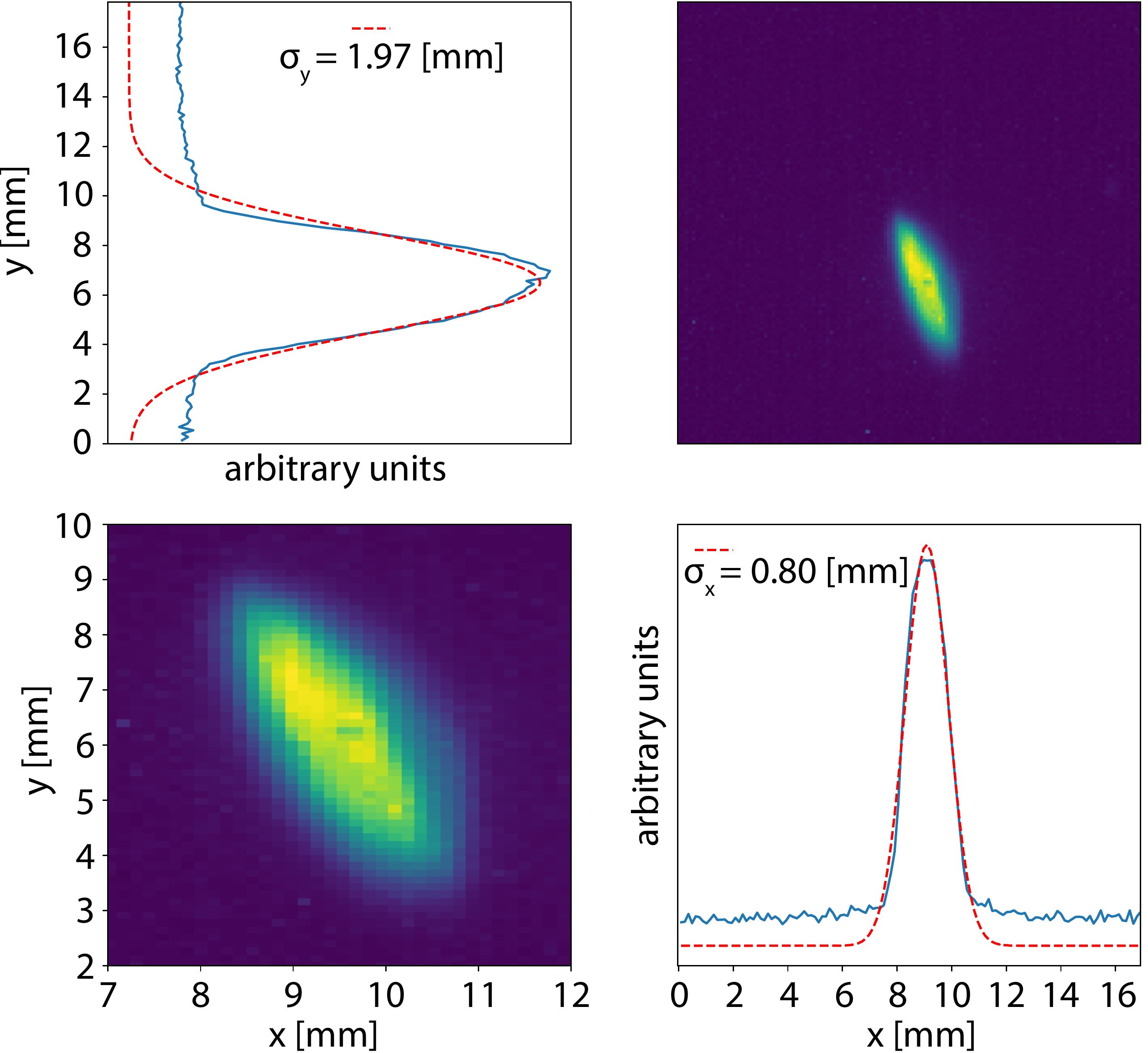}} \qquad
{\includegraphics[width=0.345\textwidth]{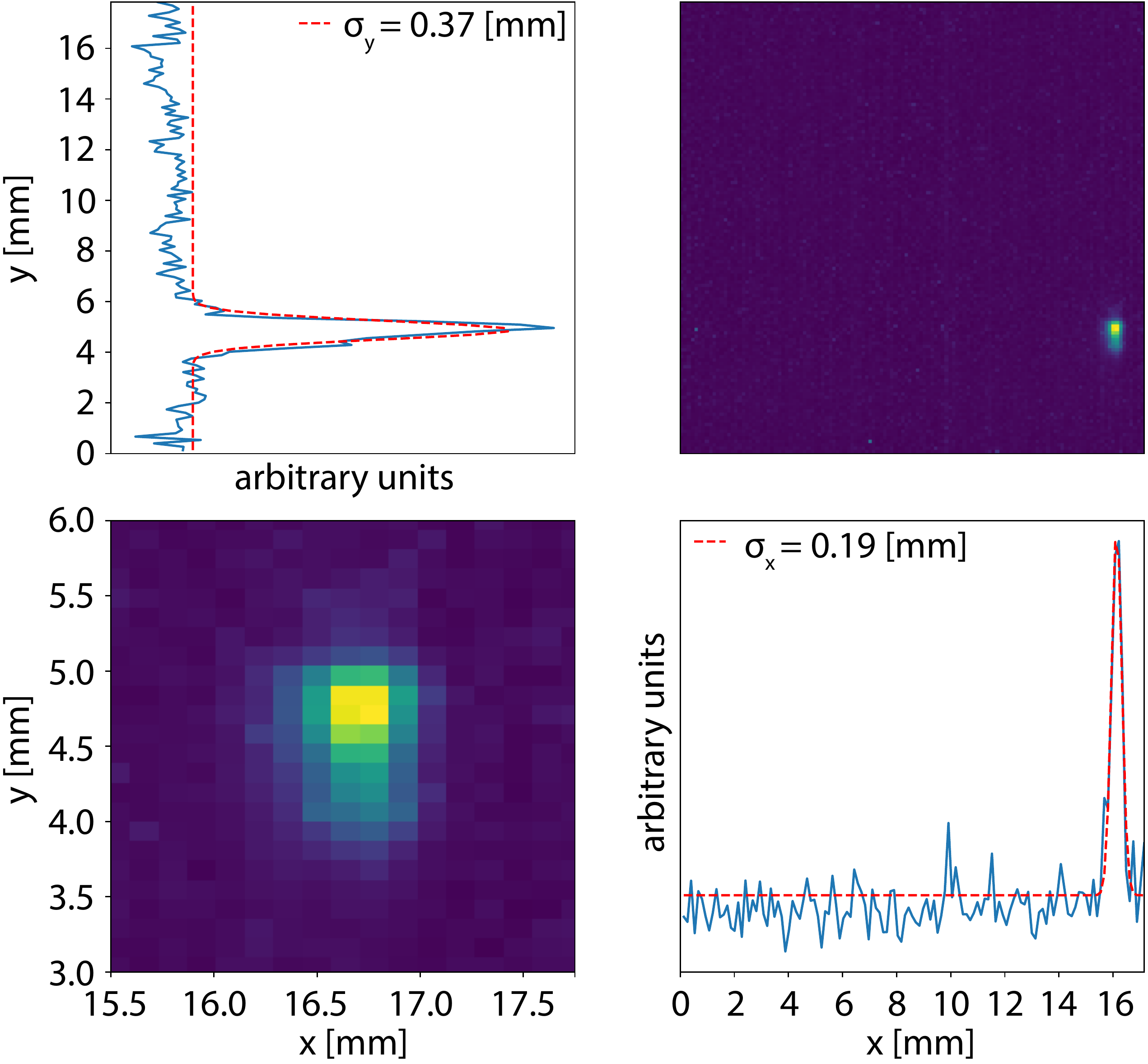}}
\caption{Initial beam shown on the left, final beam to which the algorithm converged is shown on the right.}
\label{fig:CERN_emittance}
\end{figure}

\vfill
\pagebreak
%%%%%%%%%%%%%%%%%%%%%%%%%%%%%%%%%%%%%%%%%%%%%%%%%%%%%%%%%%%%%%%%%%%%%%%%%%%%%%%%%
%%%%%%%%%%%%%%%%%%%%%%%%%%%%%%%%%%%%%%%%%%%%%%%%%%%%%%%%%%%%%%%%%%%%%%%%%%%%%%%%%
%%%%%%%%%%%%%%%%%%%%%%%%%%%%%%%%%%%%%%%%%%%%%%%%%%%%%%%%%%%%%%%%%%%%%%%%%%%%%%%%%
\section{Beam Dynamics}
%%%%%%%%%%%%%%%%%%%%%%%%%%%%%%%%%%%%%%%%%%%%%%%%%%%%%%%%%%%%%%%%%%%%%%%%%%%%%%%%%
%%%%%%%%%%%%%%%%%%%%%%%%%%%%%%%%%%%%%%%%%%%%%%%%%%%%%%%%%%%%%%%%%%%%%%%%%%%%%%%%%
%%%%%%%%%%%%%%%%%%%%%%%%%%%%%%%%%%%%%%%%%%%%%%%%%%%%%%%%%%%%%%%%%%%%%%%%%%%%%%%%%

\begin{itemize}
\item
Dynamics in high brightness, low energy electron beams is a challenge for extremely bright electron sources. For example, the LCLS-II injector is being designed to operate at a 1 MHz repetition rate with bunch parameters summarized in the table below. Such an injector capable has not yet been demonstrated \cite{ref-LCLS2injector}:
\begin{table*}[!htb]
%\caption{Experiment setup details.} % title of Table
\centering % used for centering table
\resizebox{0.6\linewidth}{!}{%
\begin{tabular}{|c|c|c|c|c|c|c|c|c|} % centered columns (4 columns)
\hline %inserts double horizontal lines
Bunch Charge [pC] & Peak Current [A] & Normalized Slice Emittance (95\%) [$\mu$m] \\ [0.2ex] % inserts table
\hline %inserts double horizontal lines
20 & 5 & 0.25  \\
100 & 10 & 0.4   \\
300 & 30 & 0.6   \\
\hline %inserts single line
\end{tabular}
}
\label{table_LCLS2}
\end{table*}

\item
Emittance improvement is extremely important and would allow for the generation of higher energy light for the same energy beam. When the emittance is too high, particles with large amplitude motion slide out of the bucket. At LCLS-II they expect to generate 12.8 keV light with an 8 GeV beam. If they can decrease the electron beam's emittance by 50\% they may be able to use the same energy electrons to create 18 keV light. High quality factor (Q) high efficiency super conducting cavities have extremely narrow bandwidths and therefore slow response times to control signals and have extremely demanding resonance control issues. For example, the LCLS-II accelerator will utilize cavities with Q $\approx 3\times10^{10}$. A pulsed machine with such high Q cavities would require advanced adaptive feed-forward iterative learning controllers to compensate for Lorentz force detuning. High current beam quality needs to be preserved over kilometer length scales despite wakefields, space charge forces, and coherent synchrotron radiation (CSR).

\item
Bunch compression is extremely important for all accelerators. Improved diagnostics for feedback-based tuning could potentially help achieve higher compression ratios while maintaining beam quality. Major beam dynamics challenges include optimal matching into bunch compressors, microbunch instability and beam halo. Detailed simulations are too slow. Although simulations have now been performed with the same number of macroparticles as particles in accelerators, they require tens of hours even when utilizing very large super computers and therefore cannot aid in online accelerator tuning and optimization \cite{ref-microbunch}:
\begin{figure}[!th]
\centering
{\includegraphics[width=0.75\textwidth]{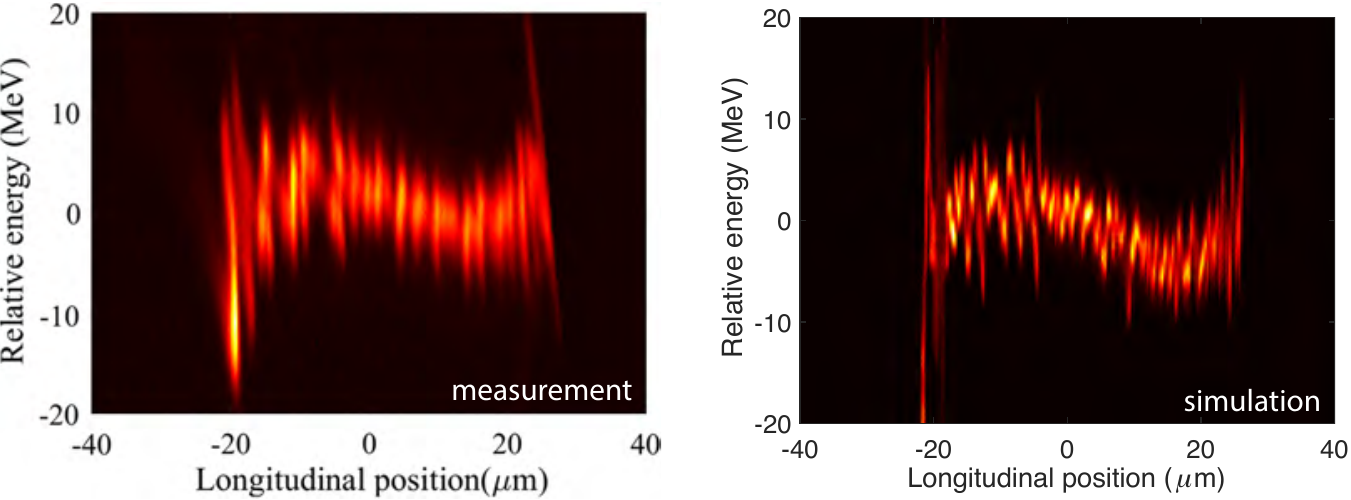}}
%\caption{x x x }
\label{fig:beam_sim}
\end{figure}

\item
Variable gap undulators add flexibility to FELs, allow for keeping beam energy constant while tuning X-ray wavelength. These devices introduce large numbers of coupled components that require advanced tuning algorithms for automated adjustment and optimization.

\item
Modern injectors are good places for developing and testing adaptive tuning techniques with few knobs and many diagnostics. Superconducting RF (SRF) systems have only local diagnostics, global tuning is difficult and must handle slow variations throughout facility, could greatly benefit from ML approaches. More precise and automated pulse-length controls are desirable for all facilities. More complex tuning procedures, such as multi-bunch/multi-color modes at FELs require advanced tuning algorithms for multiple accelerator subsystems simultaneously including the injector laser, injector, magnets, and RF systems.

\item
Plasma wakefield accelerators (PWFA) face several major challenges, the most general one is finding the an optimal approach to plasma wakefield acceleration of electron beams. There is also the unsolved problem of positron acceleration, which turns out to be much more complicated than electron acceleration. The goal is large accelerating gradients, the processes need to be efficient ($>$ 50 \%), need low energy spread ($<1$ \%), and need emittance preservation. All of these methods have extremely large parameter spaces over which it is impossible to do a brute force search. Detailed physics simulations are too slow, experiments are too few and have limited diagnostics. There are analytic models for: charge, bunch radius, bunch length, emittance, long. profile, bunch separation, and plasma density, but advanced algorithms have a role to play because all of the analytic models make simplifying and idealized assumptions. Advanced algorithms can be used to generate general longitudinal profiles for PWFA experiments. An initial guess based on theory for parameters may be used to give gaussian beam + trapezoid, then fine tuning with adaptive methods.
\begin{figure}[!th]
\centering
\includegraphics[width=0.9\textwidth]{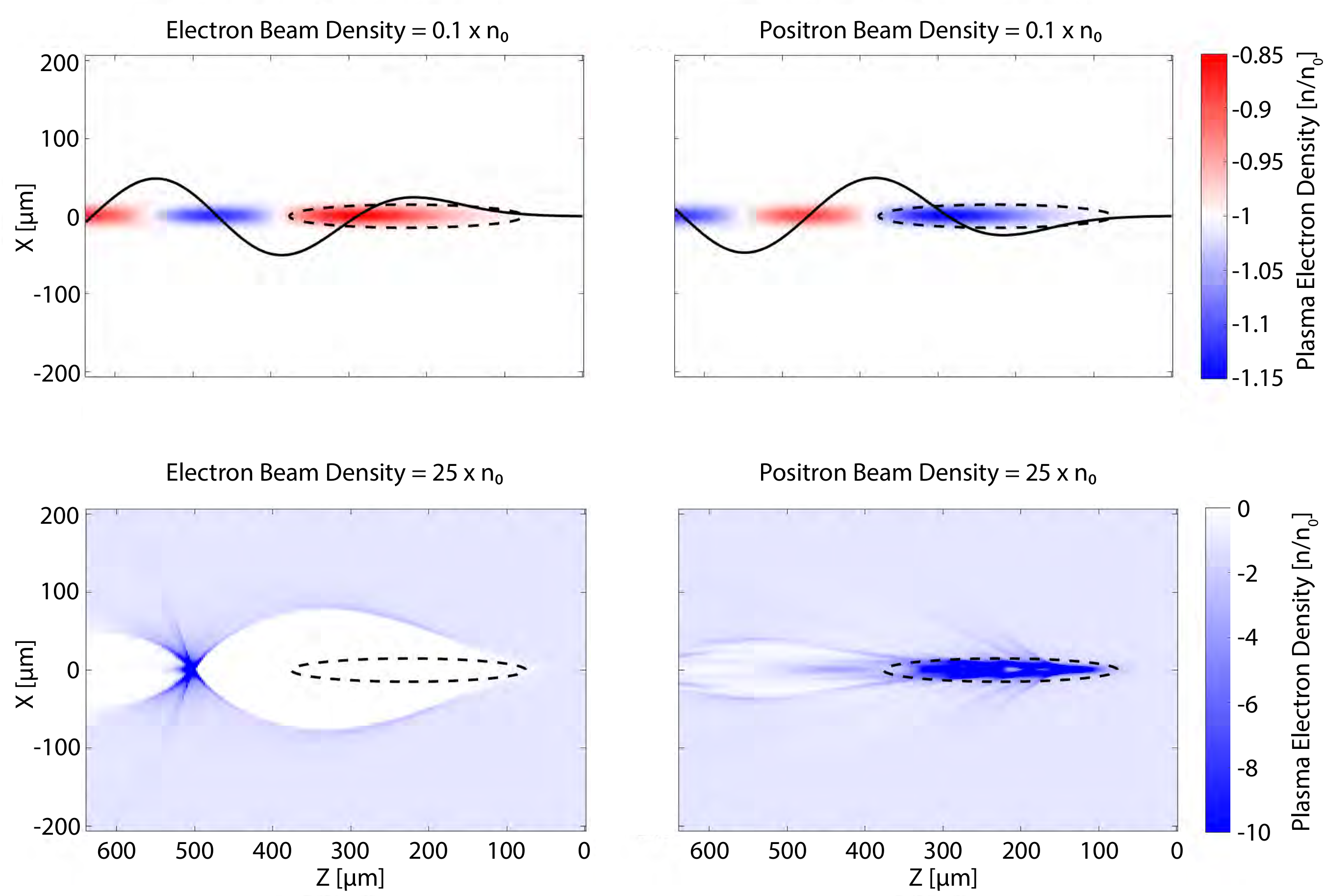}
\caption{Figure from Spencer Gessner's talk, at low space charge (top) a plasma's response to electrons is symmetric relative to its response to positrons. With higher beam charge, a large difference between accelerating electrons and positrons in a plasma becomes apparent (bottom).}
\label{fig:ep_plasma}
\end{figure}

\item
Furthermore, the analytical methods are limited because things are extremely complicated... e.g. focusing and matching conditions + preserving beam emittance (parameters are all wrapped up together). One possible solution is to just optimize on the final signal (energy gain and energy spread). Sometimes the algorithms might point us toward interesting places to look to understand the physics better (e.g. history with wakefield sims / new regimes).

\item
There is resistance to using learning and optimization algorithms to achieve core physics goals...but using optimization as a sub goal is acceptable. \textbf{Optimization gives us an opportunity to learn if the physics is missing from the model if the optimal settings do not correspond to the optimal experimental outcome.} It is essential to come up with good tests and show convincingly where these approaches work. For example, there is a claim that beam offset and emittance cannot be simultaneously optimized based on analytic description, but solutions were found using more expressive physics models. Warm start approaches are necessary to get close to expected optimum, then explore from there.

\item
Given some cost function can we run multiple PIC simulations with an optimizer to learn from the results what makes a good plasma accelerator? Use simulation + optimization to guide our understanding of why those optima work. Simulations are really slow for 3D pic, plus few opportunities to run beam (e.g. AWAKE). How to choose the next setting for the experiment? One approach is to use a very simplified model to give a rough estimate, then do detail simulations or experiments on fewer cases to zoom in on optima. Advanced algorithms have a role to play both in opt. experiments and learning underlying physics.
    
\item
The Plasma Photocathode is an advanced PWFA technique in which two gases are used with different ionization thresholds. One laser pulse then creates a plasma bubble, exciting plasma wakes in the lower threshold gas, once the bubble is excited and moving, a second laser pulse with higher energy excites electrons in the second gas, creating a plasma cathode with extremely high electric field gradients (GV/m) with the potential of creating beams with much higher 6D brightness because a relatively cool beam is born directly inside of the accelerating bubble. A beam driven version of this approach, where a driving particle bunch creates the plasma bubbles, requires extremely precise control of the 6D phase space of the driving beam. 

\item
Phase Space control is extremely important for FELs, but there are major limitations on bunch lengths for FELs. While LCLS has 10 fs resolution the goal is 100 as resolution. Soft x-rays can't be shorter than 1 fs. But hard x-rays can and with emittance spoiling can get down to 200 as. Enhanced SASE technique allows you to go to short pulses (~100as) even at soft X-rays. The basic principle is to make a very high current spike by modulating the electron beam with a laser in a wiggler then sending it through a chicane to compress. With external laser modulator you get 10s of uJ per pulse with at least 2 spikes lasing. Alternatively you can use coherent radiation generated by the tail of the electron beam to modulate the energy in the wiggler. The cool thing about that is that it's stable (modulation locked to the tail) and you can get only a single spike lasing 350 as pulse. Attosecond FELs improve peak power by 4-5 orders of magnitude compared to HHG in gas. Can do two-color as well, will upgrade XLEAP-II to do high power two-color including near TW level peak power. \textbf{Opportunity:} Challenges for tuning XLEAP - very nonlinear phase space and FEL dynamics, can we use ML models to improve the performance/design the FEL architecture (chicane R56, tune the space charge chirp etc.) for XLEAP?

\item
In general, the e-beam out of injector doesn't always have the distribution needed for a given application. Beam Shape control and phase space manipulation are required, some examples of beams that might be wanted are: flat beams, compressed beams etc. Transverse to longitudinal phase space exchange techniques exist, such as emittance exchange (EEX) for precise beam profile control and longitudinal bunch shaping \cite{ref-eex}. These techniques work by inserting masks in the beam path and utilizing quadrupoles for focusing / defocusing. There is a dedicated EEX experimental beam line at the Argonne Wakefield Accelerator (AWA). Laser shaping has also been combined with EEX for precision control of the electron longitudinal bunch shape \cite{ref-eex-prl}.
\begin{figure}[!th]
\centering
{\includegraphics[width=0.9\textwidth]{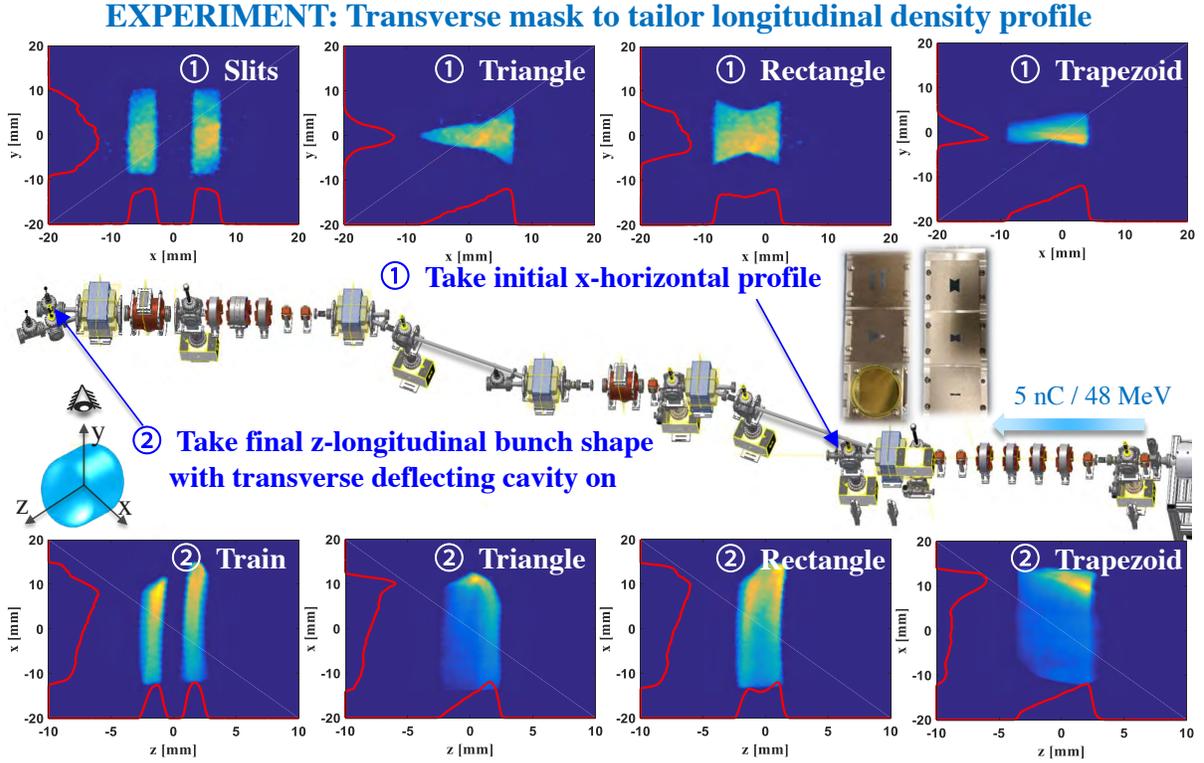}}
\caption{Figure from \cite{ref-eex-prl} on longitudinal phase space bunch control.}
\label{fig:xy}
\end{figure}

\item
Laser plasma acceleration (LPA) has its own set of unique challenges. The Bella laser plasma accelerator creates gradients of $\sim$30 GV/m. One stage of LPA has been demonstrate to create 10 GeV bunches for few pC to nC bunches with sub micron to 10 micron source size, mrad source divergence, and 1-10 micron bunch length (which depends on bubble size, such short bunches have very high current). However the energy spread of the bunches produced is $\sim$50 \% for bunches and the shot-to-shot jitter is extremely large. Creating controllable mono-energetic beams depends on on controlling and stabilizing the lasers and precisely controlling plasma density with density down-ramping. These techniques are also limited in terms of diagnostics. One method for controlling injection is by using blade in a gas jet. Moving the blade around tunes the energy of the accelerated beam and optimizes steering. It is important to measure emittance of LPA beams - can use butterfly technique. Showed that down-ramp injection gives better emittance. Butterfly diagnostic can be used as a knob to optimize beam delivery. Active plasma lenses (APL) are a useful diagnostic for beams. Radially symmetric focusing, tunable with kT/m gradient. Used in LPA staging experiment and used as a high resolution GeV magnetic spectrometer. Used APL to measure emittance of GeV class beams at BELLA $\sim$ 5 um. Liquid crystal 20nm thick based plasma mirror used to deflect PW drive laser - a good diagnostic that doesn't destroy the emittance of the beam. 
\end{itemize}

\vfill
\pagebreak
%%%%%%%%%%%%%%%%%%%%%%%%%%%%%%%%%%%%%%%%%%%%%%%%%%%%%%%%%%%%%%%%%%%%%%%%%%%%%%%%%
%%%%%%%%%%%%%%%%%%%%%%%%%%%%%%%%%%%%%%%%%%%%%%%%%%%%%%%%%%%%%%%%%%%%%%%%%%%%%%%%%
%%%%%%%%%%%%%%%%%%%%%%%%%%%%%%%%%%%%%%%%%%%%%%%%%%%%%%%%%%%%%%%%%%%%%%%%%%%%%%%%%
\section{Diagnostics}
%%%%%%%%%%%%%%%%%%%%%%%%%%%%%%%%%%%%%%%%%%%%%%%%%%%%%%%%%%%%%%%%%%%%%%%%%%%%%%%%%
%%%%%%%%%%%%%%%%%%%%%%%%%%%%%%%%%%%%%%%%%%%%%%%%%%%%%%%%%%%%%%%%%%%%%%%%%%%%%%%%%
%%%%%%%%%%%%%%%%%%%%%%%%%%%%%%%%%%%%%%%%%%%%%%%%%%%%%%%%%%%%%%%%%%%%%%%%%%%%%%%%%

\begin{itemize}

\item
Most existing non-invasive diagnostics are limited and provide only bunch-averaged information, such as beam centroids provided by beam position monitors (BPMs). For many accelerator tuning tasks, such as orbit control, BPMs are sufficient and can be used to gain other information, together with higher order mode (HOM) couplers can detect beam being off center in an RF cavity. FAST BPM measurements moving steering magnets work as expected. Work has been done to steer the beam through a section of the accelerator and find the minimum HOM signal, this tells you you're on center. Then, sending a microbunched beam deliberately off-center shows oscillation of subsequent microbunches kicked by the wakefield of the head of the beam. This method can be used to  assign where initial deflection occurred to find badly behaving beam line elements. A framing camera can image microbunches individually, this was demod at FAST, plotted centroid position of each microbunch, verified with BPM readings (these are highly correlated, plus framing camera gives much more info about what the profile of the microbunch looks like).

\item
There are also computational approaches being considered for diagnostics, one of which is to attempt to train a neural network to map accelerator parameter settings to longitudinal phase space (LPS) distributions, this approach has been demonstrated for simulations of the LCLS and FACET-II and in hardware at the LCLS \cite{ref-ML4} in which a trained NN was able to predict LPS distributions based on accelerator parameter settings, as shown in Figure \ref{fig:LCLSML}.

\item
Another approach is to adaptively tune accelerator models based on information non-invasively gathered about the beam to get an actual predictive match between model and machine. A preliminary version of such an approach was demonstrated at FACET where a non-invasive energy spread spectrum was used to adaptively tune a model in real time to give predictions of the LPS of the electron beam which was confirmed with TCAV measurements. This method was demonstrated to be extremely robust to time-variation of the system, further studies are being planned at LANSCE and FACET-II. Results from preliminary work \cite{ref-ES-FACET} are shown in Figure \ref{fig:FACET-ES}. As the simulation is adjusted to match the energy spread spectrum prediction to the measurement, it is able to track the actual LPS of the electron beam as accelerator components are changed over time, predicting single and double bunch configurations and bunch widths.
\begin{figure}[!th]
\centering
{\includegraphics[width=0.95\textwidth]{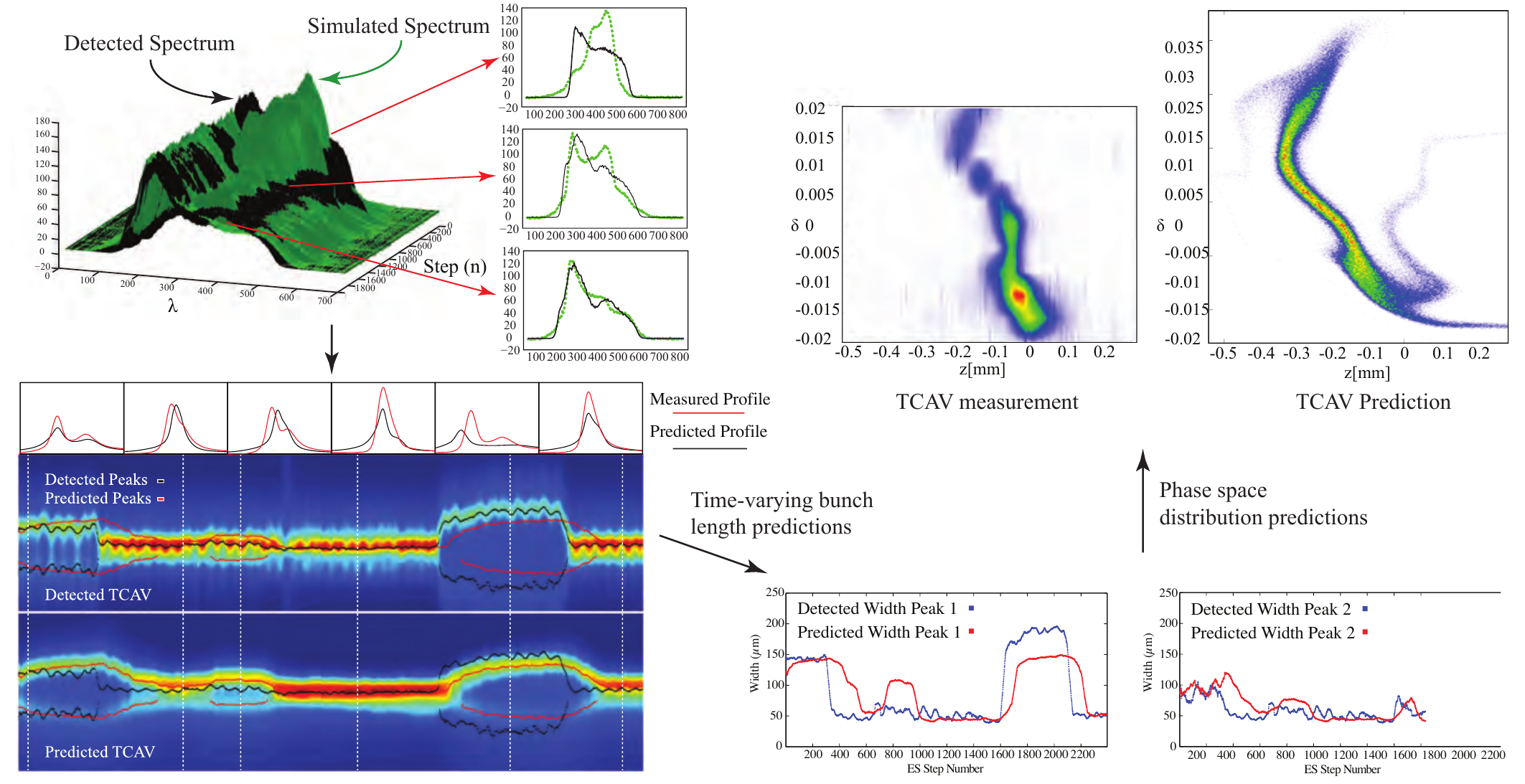}}
\caption{In this approach the adaptive tuning algorithm adjusted the parameters of a simulation until the non-destructively measured energy spread spectrum of the beam was matched by the simulation's prediction of this spectrum. Once the simulated spectrum converged to the measured spectrum the adaptive scheme was continuously run in real time while accelerator settings were adjusted and was able to track changes predicting the LPS of the beam as it changed from single to double bunch mode. Figure from \cite{ref-ES-FACET}.}
\label{fig:FACET-ES}
\end{figure}

\item
Optical Diffraction Radiation (ODR) is a non-invasive method for transverse beam profiles measurement. A beam grazes a metal edge,  a few projections are all that's needed to reconstruct. Multiple 1D $x$ profiles at different locations in the beam line are recorded to reconstruct 2D $(x,x')$ phase space distribution. Multiple diagnostics can corroborate or disagree with each other - understanding the root source of signatures can help with feedback + optimization. Bunch-to-bunch data at $<1 \ \mu$s time scales available. If there is enough charge in a micro-bunch, then if you cause the bunch to undergo a phase advance between ODR stations, you can get a reconstruction of a single bunch non-invasively.
\begin{figure}[!th]
\centering
{\includegraphics[width=0.75\textwidth]{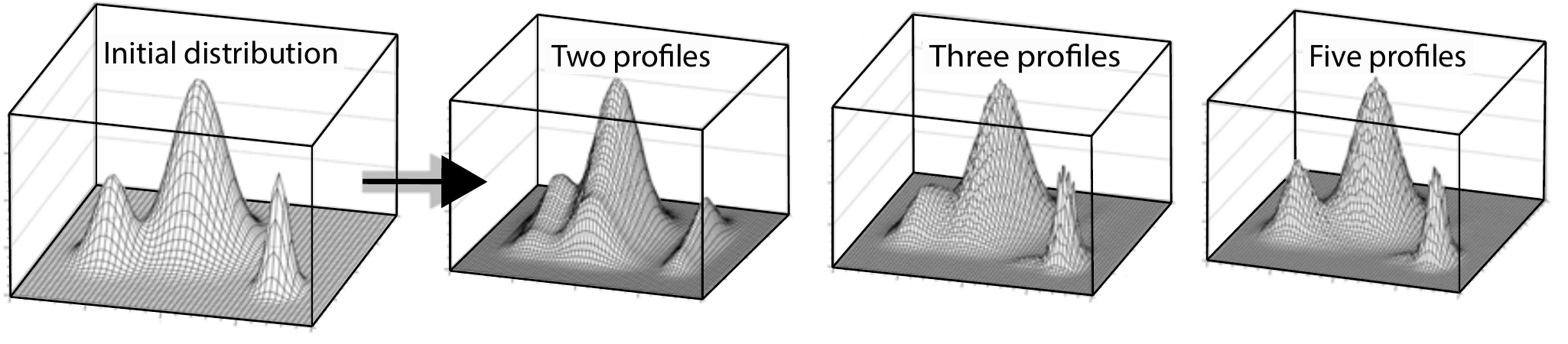}}
\caption{Figure from talk of Kip BIshofberger on application of ODR for $(x,y)$ phase space reconstruction.}
\label{fig:xy}
\end{figure}

\item
Non-invasive measurements for PWFAs. Want transverse and longitudinal measurements non-invasively shot-to-shot. PWFA Diagnostics have several issues:
\begin{itemize}
    \item General issue: PWFA has small spatial scale. Diagnostics are hard.
    \item Beam alignment: need to separate beams to measure accurately. Jitter tolerances depend on BPM uncertainty?
    \item Beam size: What is uncertainty on beam size while still preserving emittance?
    \item Beam size measurement: limited by optics. can't do much better than two microns.
    \item Butterfly emittance: use energy spread to fit emittance.
    \item Uncertainty for butterfly: consider resolution issues. No consideration of matching for this case. ``Best possible scenario" is limited by diagnostic resolution.
    \item Need to measure beam profiles... TCAV description. TCAV needs stable voltage. Resolution is 0.3 microns.
\end{itemize}
Need drive and witness beam characterization on every shot (edge radiation, butterfly emit. diagnostic). FACET: major challenge for diagnostic and control - stabilize LPS and compression against drift. LiTrack + adapt parameters to match the spectrum, also match the TCAV (non-invasive estimate of destructive TCAV diagnostic). For all adaptive methods, the convergence rate and accuracy is sensitive to initial guess. These methods will be tried at FACET-II, but collective effects may make this a challenge (plus also won't include CSR, microbunching etc). FACET-II has different challenges: want to measure LPS and stabilize compression with respect to shot-to-shot jitter of linac parameters, and measure emittance shot-to-shot. Virtual diagnostics for LPS studied in simulation with jitter estimates (use ML model based on Lucretia runs). Virtual diagnostic designs need to consider several factors:
\begin{itemize}
	\item How many shots to feed ML model to get accuracy, it took approximately 600 in a demonstrated simulation-based case (will change depending on the configuration).
	\item Need to be able to trust the virtual diagnostic, how to flag the shots we trust, also can use redundant diagnostics (need a measure of confidence).
	\item Sensitivity of model to different input diagnostics, need BC20 peak current measurement (e.g. how close to full compression).
	\item also looking at for two bunch case including smearing effect from TCAV
	\item Does TCAV image give us an answer that matches what we expect at the IP: sort of, after peak current of 30kA they don't match - need to augment with correlations with other measurements, or spectral measurements, or simulations.
	\item Want a THZ spectrometer to give us a second estimate (if difference between the two is very large, could throw out that shot).
	\item For LCLS case measured data, but also found that sometimes it does not work - sometimes the TCAV and bunch length monitor do not agree, and these give bad shots - also need to flag these measurements.
\end{itemize}
For emittance measurements, edge radiation would enable doing tuning based on emittance at multiple points along machine (simple fitting not enough to analyze image). Off-axis undulator radiation bunch length diagnostic is another possibility. The off-axis undulator radiation pattern from short electron bunches is proportional to the bunch length. Measuring the 2D radiation distribution and the integrated energy gives a measure of relative bunch length. In principle is non-destructive and can be used to measure very short bunches $\sim 1\mu m$ in length, but iterative phase reconstruction methods still need to be proved with real bunches or realistic simulations.
\begin{figure}[!th]
\centering
{\includegraphics[width=0.7\textwidth]{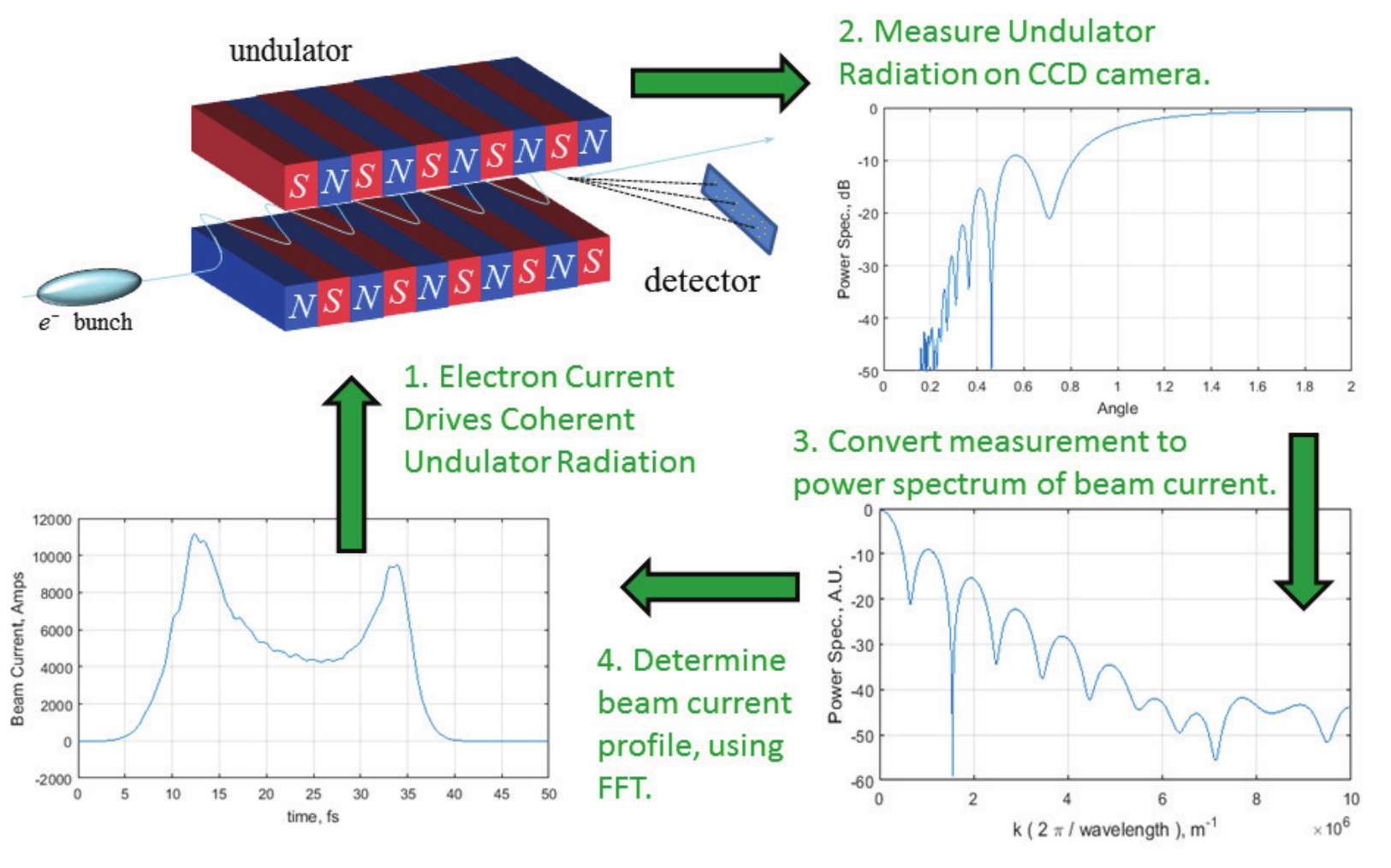}}
\caption{Figure from talk of Quinn Marksteiner on using off-axis undulator radiation for bunch profile predictions.}
\label{fig:z}
\end{figure}

\item
The ability to have autonomous accelerators is growing in importance. An autonomous accelerators: Self optimizes, self-adapts to task, finds workarounds to operating problems, makes go/no-go decisions for operating. Complexity vs consequence - accelerators are getting more complex and the consequences more serious. Example: space-borne linac for magnetotail studies - low data rates O(60) independent RF structures and can't be run from the ground (1/2 sec radio delay). This linac must operate autonomously. Medical device / food sterilization example: accelerator needs to be aware of factors (what am I sterilizing, how to efficiently set the dose, is it safe for me to run?). Accelerators for cancer therapy: Basic research needs workshop comments: doctors want more more powerful doses to better kill cancer cells (this is dangerous potentially deadly). They want improved collimation systems. Waste treatment plant example: Chicago plant processes 1.5e9 gallons/day of wastewater. Requires huge doses (14 kGy) and 500 MW of radiation power to sterilize. Compact, high power complex accelerators are here, specialists not always there to help diagnose a problem/repair a fault. Safety concerns important and have to be addressed.

\end{itemize}
\vfill
\pagebreak
%%%%%%%%%%%%%%%%%%%%%%%%%%%%%%%%%%%%%%%%%%%%%%%%%%%%%%%%%%%%%%%%%%%%%%%%%%%%%%%%%
%%%%%%%%%%%%%%%%%%%%%%%%%%%%%%%%%%%%%%%%%%%%%%%%%%%%%%%%%%%%%%%%%%%%%%%%%%%%%%%%%
%%%%%%%%%%%%%%%%%%%%%%%%%%%%%%%%%%%%%%%%%%%%%%%%%%%%%%%%%%%%%%%%%%%%%%%%%%%%%%%%%
\section{Advanced Control Methods}
%%%%%%%%%%%%%%%%%%%%%%%%%%%%%%%%%%%%%%%%%%%%%%%%%%%%%%%%%%%%%%%%%%%%%%%%%%%%%%%%%
%%%%%%%%%%%%%%%%%%%%%%%%%%%%%%%%%%%%%%%%%%%%%%%%%%%%%%%%%%%%%%%%%%%%%%%%%%%%%%%%%
%%%%%%%%%%%%%%%%%%%%%%%%%%%%%%%%%%%%%%%%%%%%%%%%%%%%%%%%%%%%%%%%%%%%%%%%%%%%%%%%%
\begin{itemize}

\item
Machine learning (ML) branches from statistics and regression. ML-based tools, such as neural networks (NN), can be trained to automatically tune and control large complex systems such as particle accelerators \cite{ref-nature,ref-ion-source,ref-ML1,ref-ML2}. ML tools are being developed to provide surrogate models to create diagnostics \cite{ref-ML3}. A NN model has been designed to predict the resonant frequency of the radio frequency quadrupole (RFQ) in the PIP-II Injector Experiment (PXIE), to be used in a model predictive control scheme \cite{ref-Edel2}. In a preliminary simulation study for a compact THz FEL, a NN control policy was trained to provide suggested machine settings to switch between desired electron beam energies while preserving the match into the undulator and a fast surrogate model was also trained from PARMELA simulation results in order to facilitate the training of the control policy \cite{ref-Edel3}. 

Powerful NN tools have also been developed for ML-based longitudinal phase space prediction of transverse deflecting cavity readings in particle accelerators, which are some of the most important diagnostics that exist for measuring a beam's longitudinal phase space \cite{ref-ML4}. A novel Bayesian optimization framework that uses sparse online Gaussian processes has been applied for quadrupole magnet tuning in an FEL \cite{ref-LCLS-Gauss,ref-ML-Adi}. Various ML tools, including clustering for identifying faulty beam position monitors (BPM) using outlier detection and ML methods for optics corrections has been developed and performed at CERN \cite{ref-CERN1,ref-CERN2,ref-CERN3,ref-CERN4,ref-CERN5}.

Although there are many ML architectures, neural networks (NN) are the most successful, in particular convolutional neural networks (CNN). The biggest lie in ML is that you sample things that are independently identically distributed, nature is not really like this, e.g. would be equivalent to saying that there is no machine drift in accelerators. NNs are optimized using (stochastic) gradient descent on your average sampled loss (error between known answers and predictions). There are many hyper-parameters that must be tuned for a NN approach:
\begin{itemize}
        \item Breaking training data down into several individual batches. Batch size: are small batches better for finding good optima? Originally batch sizes came about because it is easier to parallelize over the batch during training, but max reasonable batch size is larger than previously thought (e.g. $\sim$1000) - noise scales proportionally to $1/\sqrt{\mathrm{batchsize}}$ times learning rate (keep in mind that learning rate and batch size are correlated, so don't optimize these together during hyper-parameter tuning).
        \item Generalization error typically defined as the difference between the error on the test set and the error on the training set (you want this to be small).
        \item To get a good general model you want to sample your parameter space appropriately (i.e. not just oversample the bulk of the distribution).
        \item There is a trade-off between network complexity (depth and width) and generalization (ICLR 2016 paper on over parameterization and 2018 Belkin paper explaining how networks can generalize despite being over-parameterized).
        \item If you have a fully connected network you don't gain a lot from having deeper networks (e.g. not much difference between 3 and 10 layers). This is different if you have convolutional layers, then adding depth really helps.
        \item Very large NNs do not have local minima to be trapped in... because NNs are similar to interpolating functions (see Belkin et al arXiv 2018 and Geiger et al 2019).
        \item Infinitely wide neural nets are Gaussian Processes (Lee et al 2017 ICLR) fully connected networks reach the 'infinite' limit before CNNs.
\end{itemize}
When systems are changing with time, when there are distribution shifts one common approach is to retrain just a few of the layers of the NN on the new data, thereby using a warm start for network weights based on previously trained models. Furthermore, multiple networks, trained at different times, can be used simultaneously to give some uncertainty (check when network predictions disagree) and some may perform better than others depending on how the system has shifted. In addition to tracking mean of model ensemble, can also track the variance. Transfer learning: feature learning happens in large nets like Inception V3 trained on ImageNet -can use layers from a previously trained net and then re-train only on last layers to transfer to problems with smaller training set sizes. Robustness: models often do poorly in generalizing close to the distribution (e.g. poor performance when add noise to images), there will be classes of noise your model has not seen and it will not be robust to that. Data augmentation: very essential for improving accuracy and robustness (e.g. symmetries, random shifts and crops, cutouts / random erasing, mixup), but note some things like mixup have not been used for regression, but in some cases worth just trying for your specific case. Incorporating Symmetries: distance of test data from training manifold is important (e.g. can incorporate symmetry) - use components that commute with symmetries (equivariant) - convolutions are translation equivariant, but not scale and rotation equivariant. Incorporating other Structures (e.g. info from domain science) can be very helpful. Differentiability - can write simulations in ML languages and then can differentiate these, can also replace pieces with NNs.
\begin{figure}[!th]
\centering
{\includegraphics[width=0.6\textwidth]{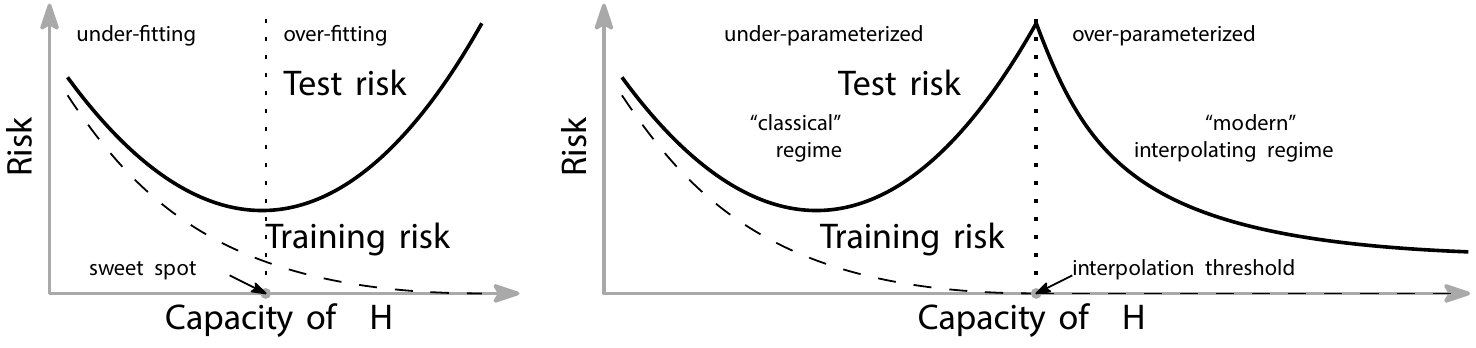}}
\caption{Figure from \cite{ref-Google} showing that unlike traditional regression approaches, once NNs become sufficiently wide and deep they can generalize and do not overfit despite such a large number of parameters.}
\label{fig:gp}
\end{figure}

\begin{figure}[!th]
\centering
{\includegraphics[width=0.85\textwidth]{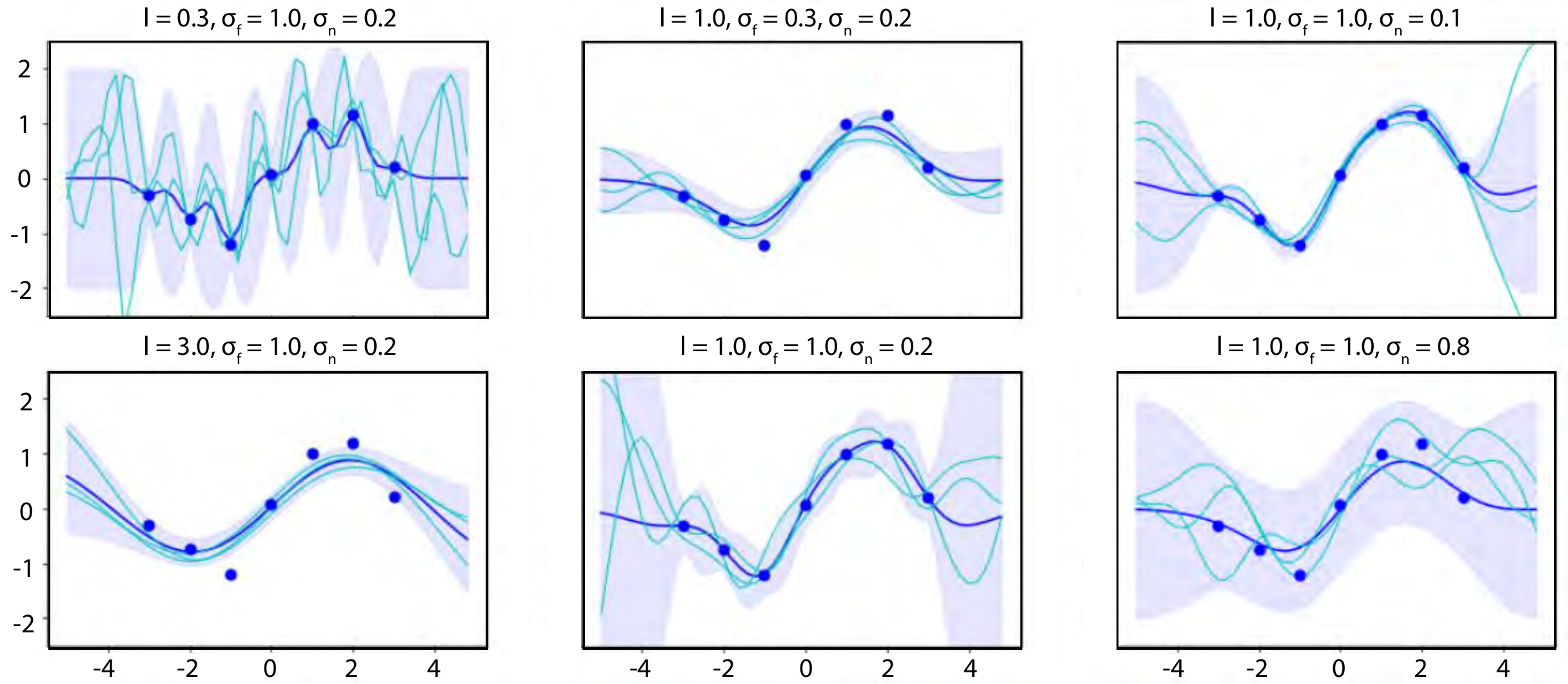}}
{\includegraphics[width=0.85\textwidth]{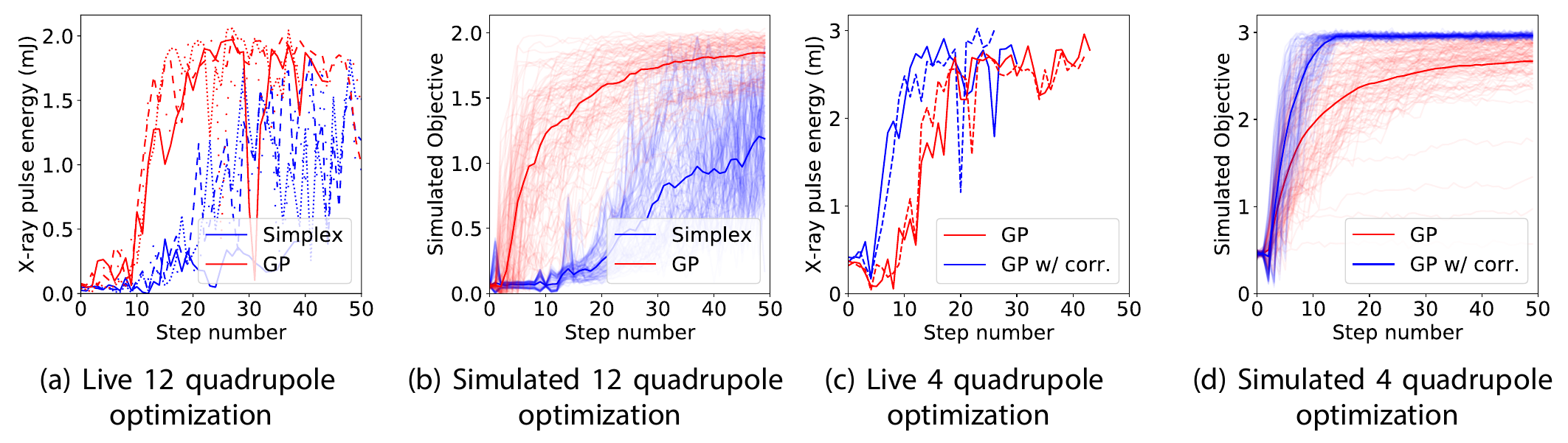}}
\caption{Figures from work being done at LCLS \cite{ref-LCLS-Gauss} and presented by Adi Hanuka at the workshop. The Gaussian Process approach is much faster than Simplex and reaches a higher maximum.}
\label{fig:gp}
\end{figure}

\begin{figure}[!th]
\centering
{\includegraphics[width=0.85\textwidth]{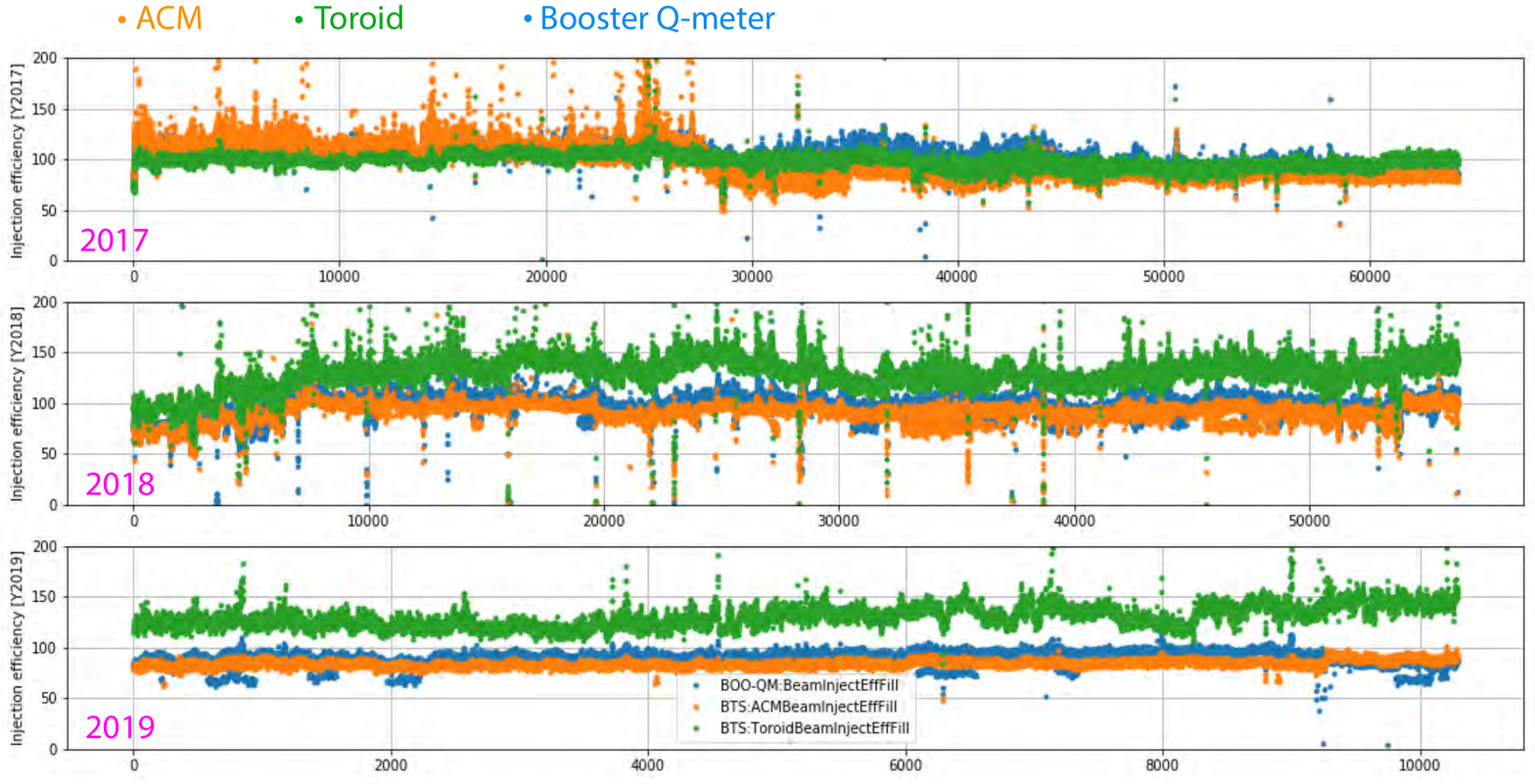}}
\caption{Figure from presentation of Faya Wang at the workshop showing three years of three different measurements of injection efficiency. This data was analyzed with ML methods to identify optimal trajectory settings as a function of temperature.}
\label{fig:spr3}
\end{figure}

\item
Large data, complex problems can benefit from automated feedback and Machine Learning (ML) approaches to detect and predict unwanted changes, extract more information from waveforms, signals, and images, improve the understanding of large, complex machines and automatically, quickly tune. For example, LCLS tuning requires $\sim$500 hours of beam time per year. Quadrupole magnets are tuned 2-5 times/day for 15 minutes at a time. Examples of ML techniques applied to accelerators:
\begin{itemize}
        \item Anomaly detection: CERN. Used clustering algorithms to find bad BPMs. ``Isolation forests", ``Local Outlier Factor", ``DBSCAN".
        \item Extract information from diagnostics: LCLS. Use CNN to correlate images with x-ray power simulations.
        \item Virtual diagnostics: FAST. Real diagnostics not always available.
        \item Virtual diagnostics: SLAC/LCLS-II/FACET-II. Correlate fast parameters with XTCAV measurments.
        \item Control and tuning: LCLS. Bayesian Optimization for FEL tuning at LCLS.
        \item Control and tuning: LCLS. ``Warm start". A trained neural network gives an initial guess for parameter settings to achieve a desired LPS and then a local model-independent feedback, such as extremum seeking (ES) tunes based on real time machine data and performs feedback to continuously optimize the system despite uncertainty and time variation.
        \item Accurate machine models: LANSCE. GPU accelerated PARMELA.
        \item Online modeling: UCLA/Pegasus. Found new beam distributions.
\end{itemize}

\item
Online FEL tuning based on Bayesian optimization not only automatically adjusts parameters, but also gives uncertainty measures on prescribed settings. Bayesian optimization builds a probabilistic model and chooses sample points in parameter space in a way that minimizes uncertainty. One very powerful Bayesian approach is based on Gaussian Processes (GP), in which a kernel function is used, typically of the form:
\begin{equation}
	k(x,x') = \sigma^2_f \exp \left ( -\frac{1}{2}(x-x')^T\Sigma(x-x') \right ) + \sigma^2_n\delta(x-x'),
\end{equation}
where $\sigma_f$ controls the amplitude of fluctuations of the unknown function being modeled, $\Sigma$ is a matrix of correlation and length scales which determines both the smoothness and coupling between separate components, and $\sigma_n$ is the standard deviation of the measured noise in the experiment. GPs combine experience/data with fast decision making, and have been shown to perform much faster than simplex methods in work being done at LCLS. The figure below shows the difference in predictions based on the same 7 data sample points, given varying levels of noise, $\sigma_n$, amplitudes, $\sigma_f$, and length scales, $l$, which are arguments of the matrix $\Sigma$ and also the results of applying GPs at LCLS from \cite{ref-LCLS-Gauss}, as shown in Figure \ref{fig:gp}.

\item 
Individual accelerators and companies are starting to develop ML tools that can be shared across the community, such as OCELOT. Radiasoft is also working on several ML projects, providing ``Sirepo," an online platform for doing accelerator physics simulations. Can use ML through their Jupyter servers and reference accelerator simulations. For example, multi-slit emittance VD at FAST -- updating with measured data (sim / training done on radiasoft servers). Optimization of Thermionic Energy Converters - want to use ML surrogate models for optimization. Web-based toolkit for ML for accelerators (link to accelerator simulations + control systems including ACNET and EPICS) -- data import/visualization -- testing out in collaboration with fermilab (modular surrogate modeling + anomaly detection).  Anomaly detection -- model-based and clustering based -- testing out at FNAL for beam loss monitors. Web-based toolbox to test accelerator control algorithms, directly interface with EPICS, improving data analytics tools for operators (dummy epics server running at radiasoft right now).

\item
ML example use case at LCLS: SPEAR3 Injection Efficiency. At SPEAR3 the injection efficiency varies over time, going up and down by as much as 15\%, and it is unclear why. The trajectory is frequently re-optimized, it would be good to understand what drives trajectory changes. A look into historic data was taken to try to find out. The data set included all trajectory knobs, BL05 gap and phase, vacuum undulator gap, outdoor ambient air temperature and ground temperature (data from 2017, 2018, 2019) - 60k data points in total, FY19 ~10k points. Used three ways of measuring the injection efficiency (some noisier, some with calibration drift). High frequency jitter in Q meter reading -- processing to remove some of the major jitter. Analyze data by year as well before applying ML (e.g. look at trajectory target during different years). NN model - related control parameters and environmental parameters to injection efficiency (then tried to examine sensitivity). Once NN model was trained, it was possible to take partial derivatives of the NN output with respect to input parameters and found the largest sensitivity to ground temperature. In each temperature zone ($\pm$ 0.5 deg C) they found the top 10 \% injection efficiency and the corresponding ideal orbits.

\begin{figure}[!th]
\centering
{\includegraphics[width=1.0\textwidth]{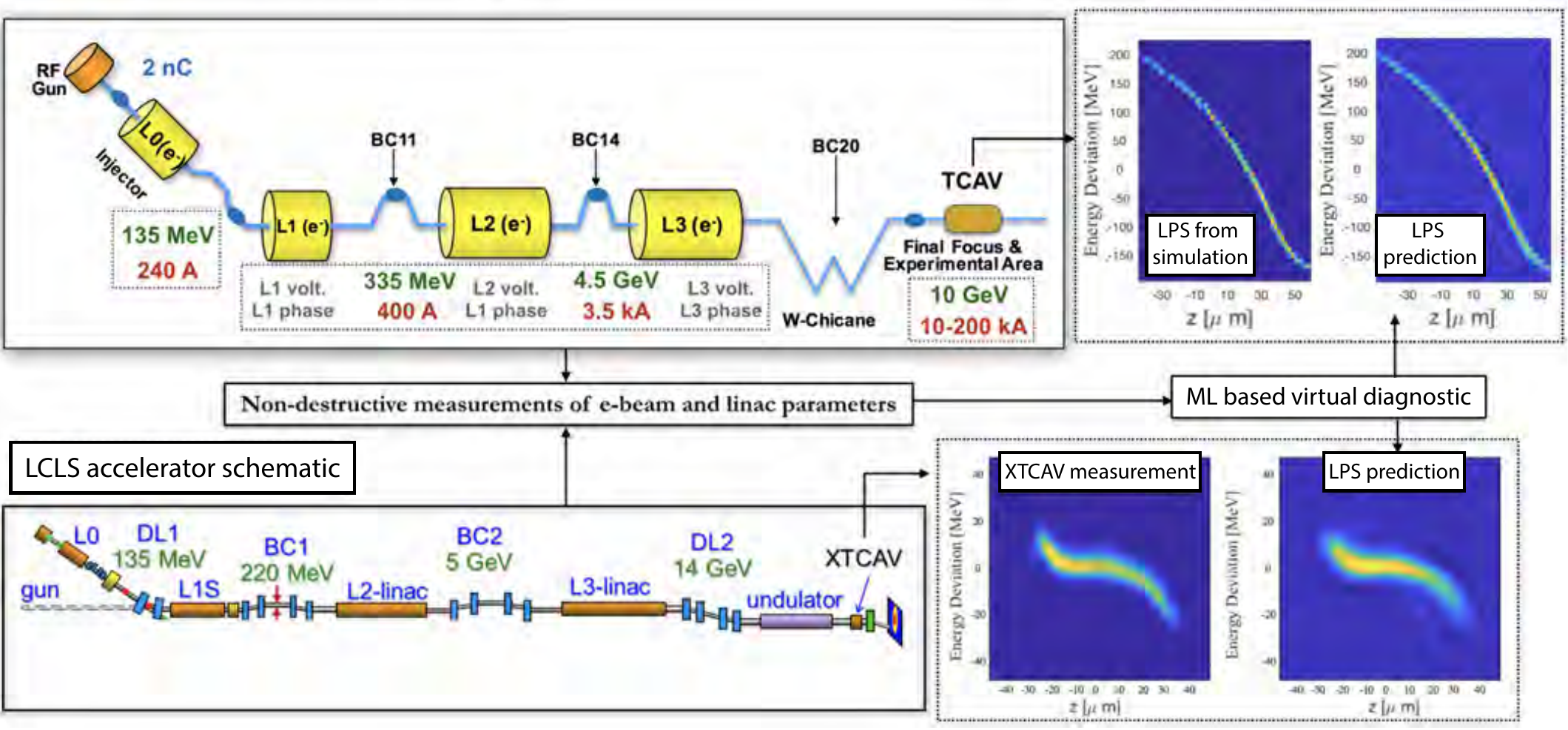}}
\caption{Figure from \cite{ref-ML4} showing NN-based predictions of the LPS of FACET-II, compared to simulations, and of LCLS, compared to TCAV measurements.}
\label{fig:LCLSML}
\end{figure}

\item
Model validation, optimization and uncertainty quantification using the Mystic package. Just using most ML methods, such as NNs, it is fundamental difficulty for ML to produce rigorously validated models of complex physics. Difficult to decide ``How good is my ML model?" Usually we decide this post-learning but you would like to incorporate this in learning process. Would like to do this as a game-theoretic optimization. It is possible to use properly chosen kernels to transfer nonlinear data into higher dimensional space and linearize your problem. Important to also use physics constraints to make penalty functions for your optimization. {\it mystic} has packages that easily incorporate many constraints for optimization, including Lipschitz cones in hyperspace and the ability to generate predictors that minimize the likelihood of misclassification. One example of {\it mystic} performing non-convex global optimization is shown in Figure \ref{fig:UQ}.

\begin{figure}[!th]
\centering
{\includegraphics[width=1.0\textwidth]{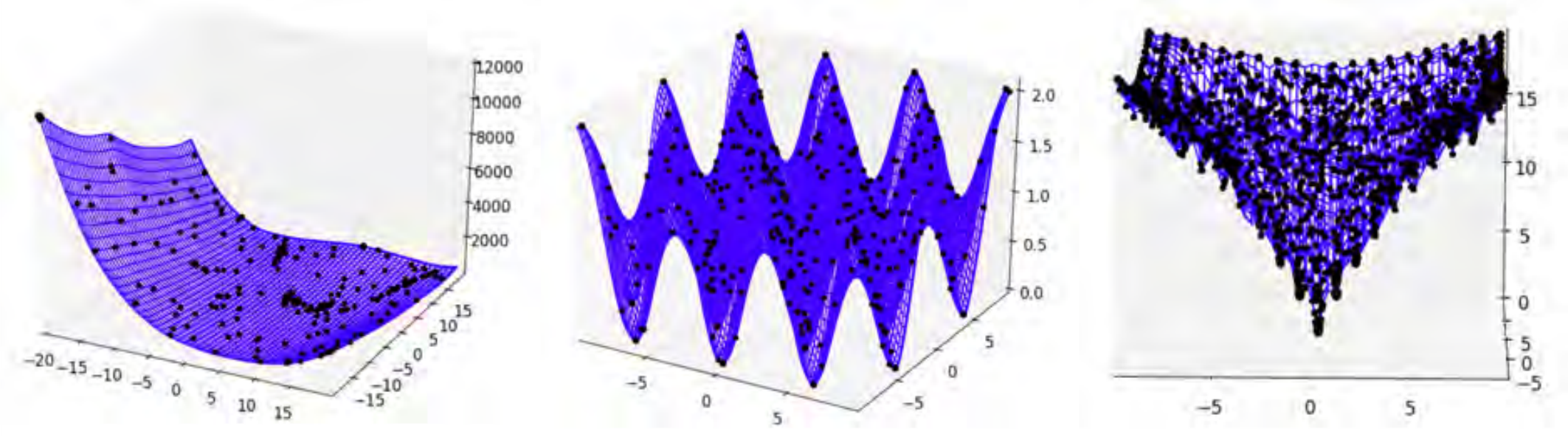}}
\caption{The {\it mystic} framework provides way to incorporate physics-based constraints for non-covex global optimization.}
\label{fig:UQ}
\end{figure}

\item
ML example use case at CEBAF: CEBAF - first large high power SRF machine in US, over 25 years operation. Cavities strongly coupled, when one shakes they all shake. Operators don't know which cavity caused the trip so they just reduce the gradient. To keep trips below 10/hr they have to run below design energy, mostly due to microphonics and field emission. There was a need to automate fault detection and classification. Different trips are accompanied by different features, 7 different types identified. Questions: which cavity caused trip, what type of trip, what should we do? Used deep learning model and got good results for prediction. Want to use deep learning models to not only predict but prevent trips in future.

\begin{figure}[!th]
\centering
{\includegraphics[width=0.85\textwidth]{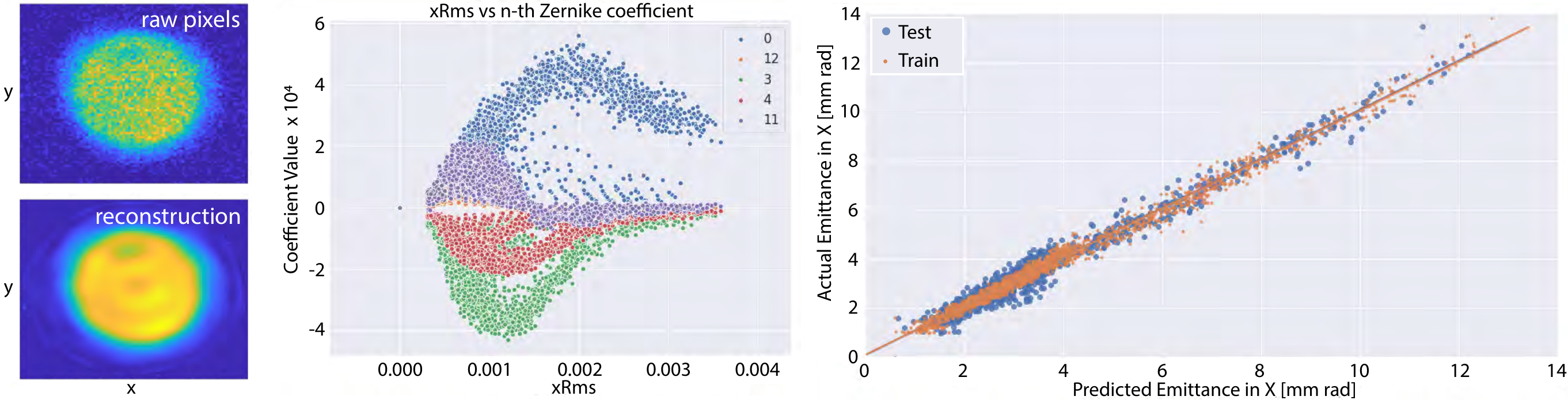}}
\caption{Figure from talk of Jerry Ling on application of Zernike polynomials and NNs for emittance measurements.}
\label{fig:cern}
\end{figure}

\begin{figure}[!th]
\centering
{\includegraphics[width=0.85\textwidth]{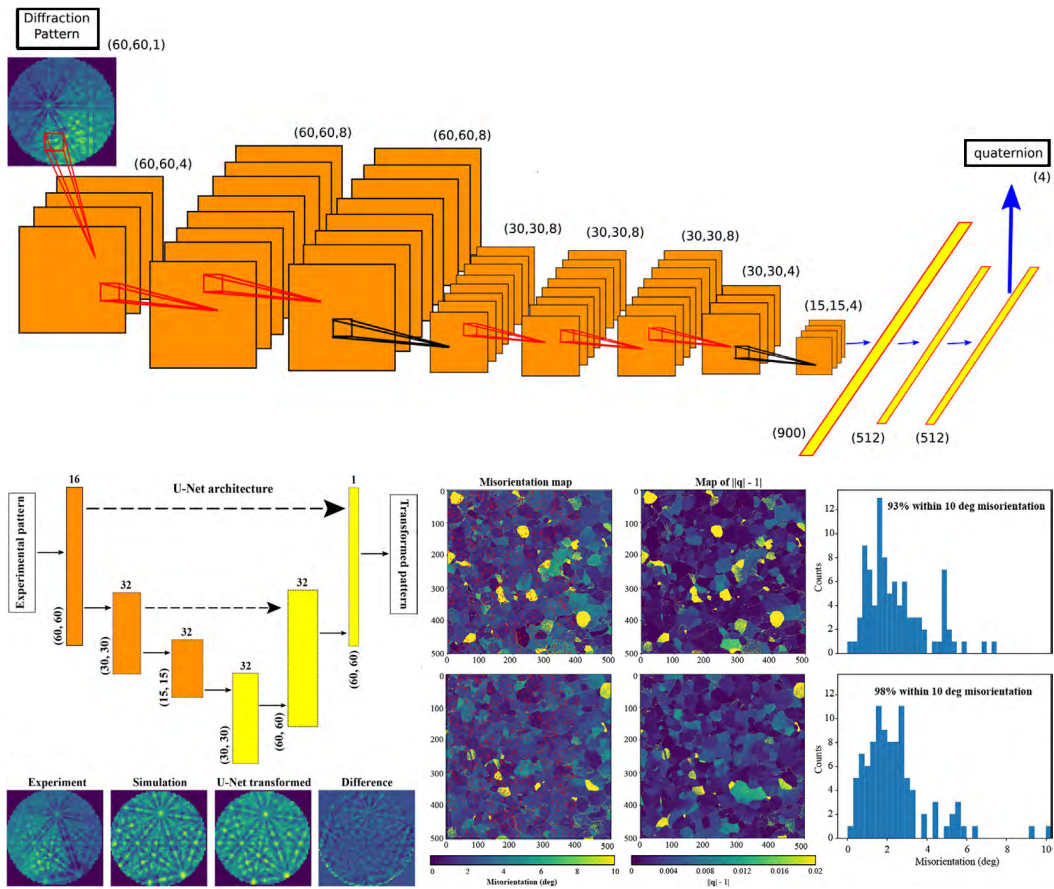}}
\caption{U-net is trained to map experimental patterns to simulated patterns to be input into a simulation trained CNN. Another approach is to re-train just the last three layers of the simulation-trained CNN using $\sim$1 K experimental measurements. Both approaches predicted $>93$\% of orientations within $<$10 deg error on experimentally collected data. Figures from \cite{ref-ML-retrain}.}
\label{fig:Unet}
\end{figure}

\item
ML example use case at CERN: CERN has many coupled accelerators (beam prepared in one and sent to another) operating simultaneously and requires scheduled beam delivery to each from injectors. A supercycle is a communal schedule across the injector complex. The SPS, feeding the LHC is top priority, smaller injectors fill space in between. A supercycle has some requirements: each user is given a min num of shots/unit time, each experiment has minimum time between shots, magnet switching constraints. Currently a supercycle is built manually. Sometimes changed many times during MD (90 times/day record). Optimizing this gives users more data and saves money. They have created an optimization algorithm that gives the most beam to everyone. The algorithm is brute force and uses topological equivalence to be complete without being exhaustive. Algorithm uses a method called local 'signposting' to map the permutation space of decision points.

\item
ML example use case at SLAC: LCLS and FACET-II at LCLS are developing a neural network (NN) approach to predicting the longitudinal phase space (LPS) of the electron beam by training the NN to relate machine parameter settings to recorded LPS images, as shown in the results from \cite{ref-ML4} in Figure \ref{fig:LCLSML}.

\item
ML example use case at CERN: at CERN there is work ongoing to predict measure beam emittance based on machine component settings, both using Zernike polynomial representations and a neural network. While both methods have had good results with simulation data, the NN approach is more accurate, preliminary results are shown in Figure \ref{fig:cern}.

\item
Retraining and domain transfer for CNN methods: Recently, a novel ML approach to the mapping of diffraction images to crystal orientations for electron backscatter diffraction (EBSD) was demonstrated which sped up the reconstruction process by several orders of magnitude \cite{ref-ML-retrain}. Because EBSD data is relatively expensive (time consuming, hours per single sample scan) to collect, the researchers started by creating hundreds of thousands (300 K) of diffraction pattern simulations. They then used pairs of diffraction patterns and their corresponding crystal orientations to train a deep CNN to automatically map diffraction patterns to orientations. Once the CNN was trained they tested 2 approaches to make the CNN useful for actual experimental data without having to rely on extremely large experimental data sets: 1). For the domain transfer approach they created a U-net, a smaller neural network which was trained to map experimentally collected diffraction images to their simulated counterparts, the training of this smaller network required only $\sim$1 K experimental diffraction patterns (compared to 300 K for the CNN). This approach allowed them to take raw measured diffraction images, map them to simulated counterparts, and then feed them into the CNN to accurately predict their crystal orientation. 2). In the retraining approach they again utilized only $\sim$1 K measured diffraction patterns to re-train only the last few layers of the CNN and it could then accurately map experimentally collected diffraction patterns to their corresponding crystal orientations. Such an approach has great potential to be used for particle accelerators for either direction: (A) Training a NN using simulation data if it is easy to generate and then re-training on a few actual accelerator measurements or (B) Training a NN using experimentally collected accelerator data for computationally expensive processes (space-charge dominated beams, CSR, etc...) and then applying re-training to match it to particular complex simulations that are very computationally expensive to create new, faster, more accurate versions of these simulations.

\item
Adaptive machine learning: To supplement ML approaches, a local, model-independent feedback algorithm exists that can optimize and tune noisy complex systems and adapt to time-varying features and distributions \cite{ref-ES,ref-Sch-Sch,ref-ES-book,ref-Sch-Sch-2}. This approach has been demonstrated for RF buncher cavity phase control \cite{ref-ES-LANSCE-phase} and resonance control \cite{ref-ES-LANSCE} at the LANSCE proton linac, for minimizing betatron oscillations in a time-varying lattice in the SPEAR3 synchrotron \cite{ref-ES-SPEAR3}, as a non-invasive diagnostic for FACET \cite{ref-ES-FACET}, and for output power maximization in both the LCLS and EuXFEL FELs \cite{ref-ES-EuXFEL}. Recently an adaptive ML approach has been developed in which a NN was trained to map longitudinal phase space distributions to the accelerator parameters required to achieve them \cite{ref-ES-NN}. The NN alone would not work because the system was drifting with time, but it did give a good approximation of the required settings in the global parameter space after which a local model-independent extremum seeking method is able to zoom in on and track the optimal time-varying parameter settings, as shown in Figure \ref{fig:warm_start}.
\begin{figure}[!th]
\centering{
\includegraphics[width=0.5\textwidth]{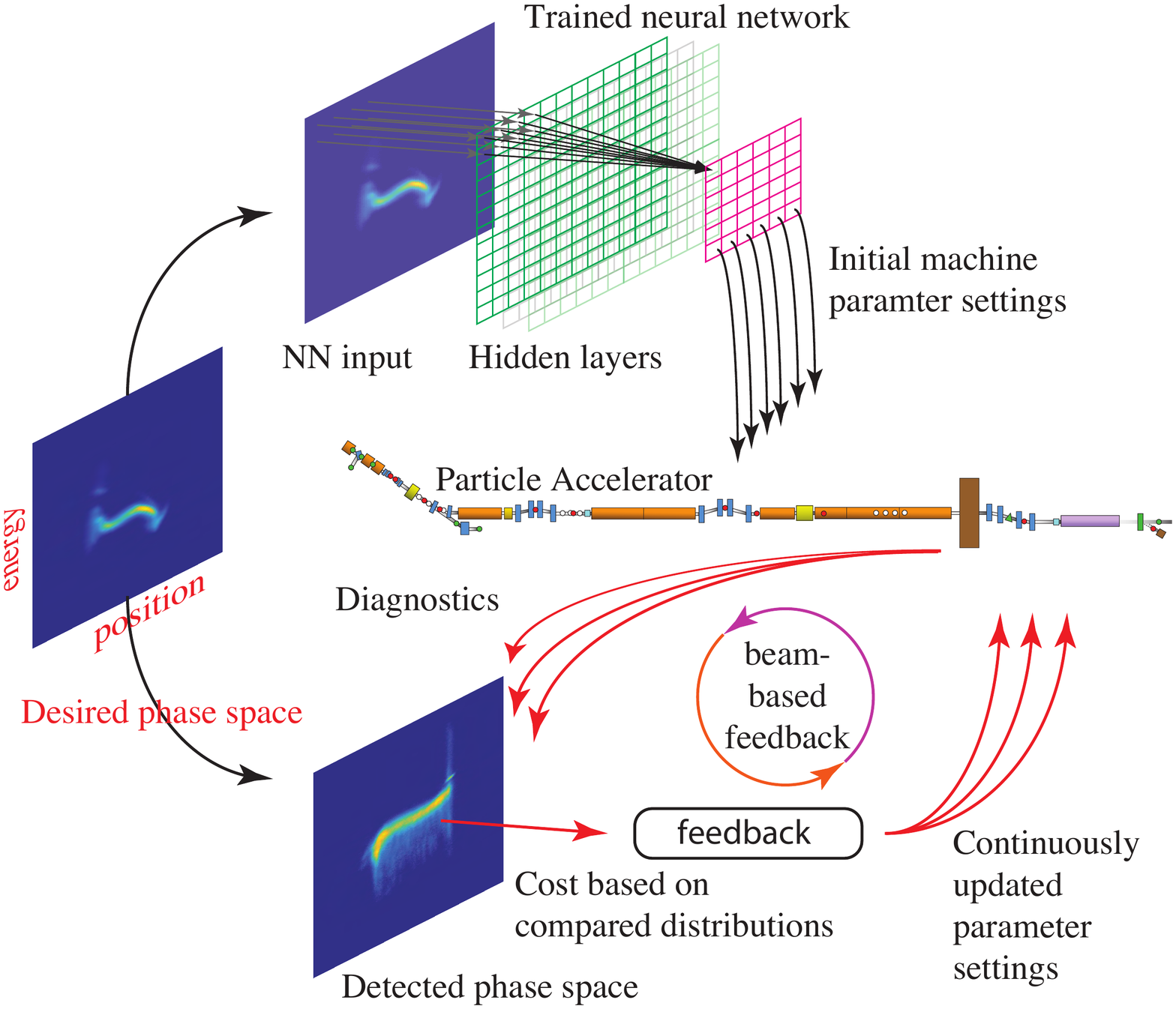}
\includegraphics[width=1.0\textwidth]{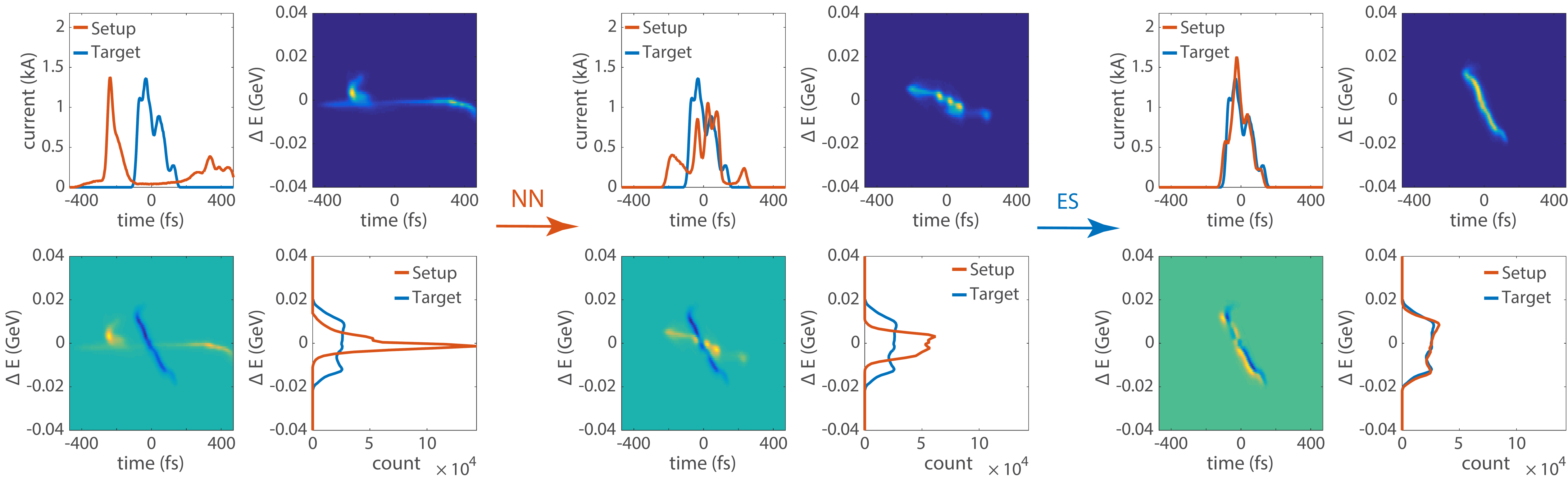}
}
\caption{{\bf Top:} Overview of the adaptive ML approach in which a NN gives an initial guess and then adaptive feedback zooms in on and tracks optimal settings while both the initial beam conditions and accelerator components drift with time. {\bf Bottom:} Experimental results at the LCLS FEL showing the initial guess of the NN after which adaptive feedback was able to zoom in on the optimal solution.}
\label{fig:warm_start}
\end{figure}

\end{itemize}
\vfill
\pagebreak
%%%%%%%%%%%%%%%%%%%%%%%%%%%%%%%%%%%%%%%%%%%%%%%%%%%%%%%%%%%%%%%%%%%%%%%%%%%%%%%%%%%%%%%%%%%%%%%%%%%%%%%%%%%%%%%%%%%%%%%%%%%%%%%%%%%%%%%%%%%%%%%%%%%%%%%%%%%%%%%%%%%%%%%%%%%%%%%%%%%%%%%%%%%%%%%%%%%%%%%%%%%%%%%%%%%%%%%%%%%%%%%%%%%%%%%%%%%%%%%%%%%%%%%%%%%%

\section{List of Acronyms}
\begin{eqnarray}
\mathrm{FEL} &&  \qquad \mathrm{Free\ Electron \ Laser} \nonumber \\
\mathrm{linac} && \qquad \mathrm{Linear \ Accelerator} \nonumber \\
\mathrm{LCLS} && \qquad \mathrm{Linac \ Coherent \ Light \ Source} \nonumber \\
\mathrm{XFEL} && \qquad \mathrm{X-ray \ FEL} \nonumber \\
\mathrm{EuXFEL} && \qquad \mathrm{European \ X-ray \ FEL} \nonumber \\
\mathrm{LANSCE} && \qquad \mathrm{Los \ Alamos \ Neutron \ Science \ Center} \nonumber \\
\mathrm{AWAKE} && \qquad \mathrm{Advanced \ Proton \ Driven \ Wakefield \ Acceleration \ Experiment} \nonumber \\
\mathrm{FACET}  && \qquad \mathrm{Facility \ for \ Advanced \ Accelerator \ Experimental Tests} \nonumber \\
\mathrm{MARIE} && \qquad \mathrm{Matter \ Radiation \ Interactions \ in \ Extremes} \nonumber \\
\mathrm{TCAV} && \qquad \mathrm{Transverse \ Deflecting  \ Cavity} \nonumber \\
\mathrm{LPS} && \qquad \mathrm{Longitudinal \ Phase \ Space} \nonumber \\
\mathrm{fs} && \qquad \mathrm{femtosecond} \ = \ 10^{-15} \ \mathrm{seconds} \nonumber \\
\mathrm{SASE} && \qquad \mathrm{Self-Amplified \ Spontaneous \ Emission} \nonumber \\
\mathrm{PWFA} && \qquad \mathrm{Plasma \ Wakefield \ Accelerator} \nonumber \\
\mathrm{LPA} && \qquad \mathrm{Laser \ Plasma \ Acceleration} \nonumber \\
\mathrm{ML} && \qquad \mathrm{Machine \ Learning} \nonumber \\
\mathrm{NN} && \qquad \mathrm{Neural \ Network} \nonumber \\
\mathrm{CNN} && \qquad \mathrm{Convolutional \ Neural \ Network} \nonumber \\
\mathrm{ES} && \qquad \mathrm{Extremum \ Seeking} \nonumber \\
\mathrm{CSR} && \qquad \mathrm{Cpherent \ Synchrotron \ Radiation} \nonumber \\
\mathrm{RF} && \qquad \mathrm{Radio \ Frequency} \nonumber \\
\mathrm{SRF} && \qquad \mathrm{Superconducting \ RF} \nonumber \\
\mathrm{EEX} && \qquad \mathrm{Emittance \ Exchanger} \nonumber \\
\mathrm{AWA} && \qquad \mathrm{Argonne \ Wakefield \ Accelerator} \nonumber \\
\mathrm{APL} && \qquad \mathrm{Active \ Plasma \ Lense} \nonumber \\
\mathrm{BPM} && \qquad \mathrm{Beam \ Position \ Monitor} \nonumber \\
\mathrm{HOM} && \qquad \mathrm{Higher \ Order \ Mode} \nonumber \\
\mathrm{ODR} && \qquad \mathrm{Optical \ Diffraction \ Radiation} \nonumber
\end{eqnarray}

\vfill
\pagebreak
%%%%%%%%%%%%%%%%%%%%%%%%%%%%%%%%%%%%%%%%%%%%%%%%%%%%%%%%%%%%%%%%%%%%%%%%%%%%%%%%%%%%%%%%%%%%%%%%%%%%%%%%%%%%%%%%%%%%%%%%%%%%%%%%%%%%%%%%%%%%%%%%%%%%%%%%%%%%%%%%%%%%%%%%%%%%%%%%%%%%%%%%%%%%%%%%%%%%%%%%%%%%%%%%%%%%%%%%%%%%%%%%%%%%%%%%%%%%%%%%%%%%%%%%%%%%

%%%
%%%
%%%
%%%
%%%
%%%

\end{document}